\documentclass[aps,prl,twocolumn,showpacs,showkeys,footinbib,superscriptaddress,longbibliography]{revtex4-1}

\usepackage{amsmath}
\usepackage{amssymb}
\usepackage{graphicx}
\usepackage{bm}
\usepackage{color}
\usepackage{relsize}
\usepackage{braket}
\usepackage[caption=false]{subfig}

\usepackage{hyperref}
\usepackage[all]{hypcap}

\begin{document}
\title{Resonant Tunneling Anisotropic Magnetoresistance Induced by Magnetic Proximity}
\author{Chenghao Shen}
\affiliation{Department of Physics, University at Buffalo, State University of New York, Buffalo, NY 14260, USA}
\author{Timothy Leeney}
\affiliation{Department of Physics, University at Buffalo, State University of New York, Buffalo, NY 14260, USA}
\author{Alex Matos-Abiague}
\affiliation{Department of Physics and Astronomy, Wayne State University, Detroit, MI 48201, USA}
\author{Benedikt  Scharf}
\affiliation{Department of Physics, University at Buffalo, State University of New York, Buffalo, NY 14260, USA}
\author{Jong E. Han}
\affiliation{Department of Physics, University at Buffalo, State University of New York, Buffalo, NY 14260, USA}
\author{Igor \v{Z}uti\'c}
\affiliation{Department of Physics, University at Buffalo, State University of New York, Buffalo, NY 14260, USA}
\date{\today}
\begin{abstract}
We reveal that the interplay between Rashba spin-orbit coupling and proximity-induced magnetization in a two-dimensional electron gas leads to peculiar transport properties and large anisotropy of magnetoresistance. While the related tunneling anisotropic magnetoresistance (TAMR) has been extensively studied before, we predict an effect with a different origin arising from the evolution of a resonant condition with the in-plane rotation of magnetization and having a much larger magnitude. The resonances in the tunneling emerge from a spin-parity-time symmetry of the scattering states. However, %
such a symmetry is generally absent from the system itself and only appears for certain parameter values.  Without resonant behavior in the topological surface states of a proximitized
three-dimensional topological insulator (TI), TAMR measurements can readily distinguish them from often misinterpreted trivial Rashba-like states inherent to many TIs.
\end{abstract}


\maketitle

\section{I. Introduction}

Tunneling magnetoresistance (TMR) has enabled remarkable advances in spintronic applications~\cite{Moodera1995:PRL,Parkin2004:NM,Yuasa2004:NM,
Zutic2004:RMP,Maekawa:2006}. While TMR devices require multiple ferromagnetic leads, in the presence of spin-orbit coupling (SOC) even a single ferromagnet (F) yields MR with the change of its magnetization direction. This makes the resulting tunneling anisotropic magnetoresistance 
(TAMR)~\cite{Gould2004:PRL,Moser2007:PRL,Chantis2007:PRL,MatosAbiague2009:PRBa,MatosAbiague2009:PRBb,Wang2013:PRB,Park2008:PRL,%
Uemura2009:APL,Wimmer2009:PRB,Ruster2005:PRL,Saito2005:PRL,Fabian2007:APS} a promising effect for scaled-down devices and design simplifications.  
Since the TAMR originates from the interplay between magnetization, {\bf M} and SOC, it can also be used for experimentally probing emergent phenomena from interfacial SOC fields in both normal and superconducting heterostructures~\cite{Gao2007:PRL,Park2011:NM,Hogl2015:PRL,Martinez2020:PRA,Vezin2020:PRB}.

\begin{figure}[t]
\vspace{-0.25cm}
\centering
\includegraphics*[width=9.3cm]{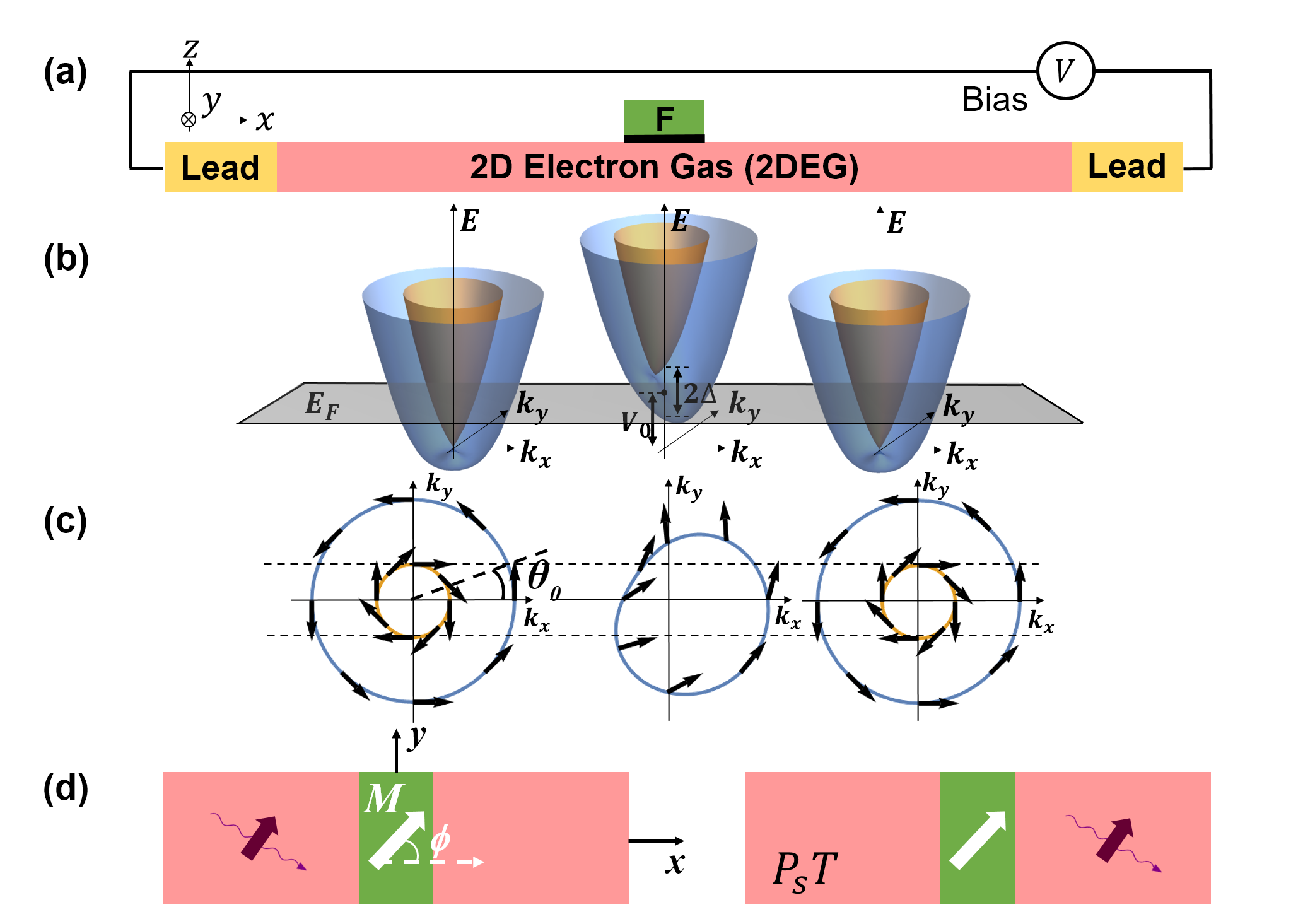}
\caption{(a) Planar geometry,  
the current flows in the 2DEG. 
(b) Band structure in the 2DEG  with the Fermi energy $E_F$ and the effective 
 barrier region (middle) of height $V_0$ and exchange field $\Delta$. 
(c) Corresponding Fermi contours, 
arrows denote the spin orientation. 
Dashed lines:
the range of a conserved wave vector $k_y$ in the scattering states. For incident angles exceeding 
$\theta_0$ backscattering is suppressed. (d) Action of the $\mathcal{P}_{s}\mathcal{T}=\mathcal{P}\sigma_z\mathcal{T}$ 
 operator on an incident wave with an in-plane spin transforms the incident wave on the left side of the barrier (left panel) into itself but as a transmitted wave on the right side of the barrier (right panel). The magnetization orientation 
${\mathbf M}$, defined by the in-plane polar angle $\phi$.}
\label{fig:setup}
\end{figure}

In common TAMR devices, a ferromagnetic lead generates spin polarization $P$, which together with the SOC strength determines the transport anisotropy. In vertical tunneling devices the in-plane TAMR is rather small (typically $< 1$ \%) even for large $P$~\cite{Gould2004:PRL} and 
exchange energies $>$ eV~\cite{Moser2007:PRL,Park2008:PRL,Uemura2009:APL}. In this work we propose a novel geometry in which a spin-polarized 
lead~\cite{Zutic2004:RMP,Egues2003:APL} is not 
required, but the TAMR is enhanced. We consider tunneling through a gated barrier with proximity-induced magnetism from F on top of a two-dimensional electron gas (2DEG),  shown in Fig.~\ref{fig:setup}(a). Magnetic proximity effects offer a versatile method  to transform materials with measured proximity-induced exchange energies up to 
tens of meV~\cite{Zutic2019:MT} and extending over 10 nm~\cite{Akimov2017:PRB,Korenev2019:NC,Takiguchi2019:NP19,Betthausen2012:S}. 

In the presence of Rashba SOC~\cite{Zutic2004:RMP} the transport through the magnetic barrier becomes anisotropic with respect to the magnetization ${\mathbf M}$. 
Remarkably, the predicted in-plane TAMR is one to two orders of magnitude larger than in most TAMR vertical devices. Surprisingly, this is realized in the proposed device with nonmagnetic leads and the proximity-induced exchange splitting typically two orders of magnitude smaller than the exchange energy in the ferromagnetic lead of other TAMR devices. We find that the enhanced MR sensitivity to changes in ${\mathbf M}$, even for small exchange fields, originates from the emergence of a spin-parity-time 
($\mathcal{P}_{s}\mathcal{T}$) symmetry of the scattering states, where $\mathcal{T}$ is the time reversal and $\mathcal{P}_s$ the inversion of both position and spin, generalizing the well-known PT-symmetry~\cite{Bender1999:JOMP,El-ganainy2018:NP}. The emergence of  the $\mathcal{P}_{s}\mathcal{T}$ symmetry leads to resonances in the transmission, which are highly sensitive to 
${\mathbf M}$ and result in an enhanced TAMR.

As illustrated in Figs.~\ref{fig:setup}(b) and ~\ref{fig:setup}(c), in the magnetic barrier region the Fermi contour of the 2DEG states is shifted perpendicular to ${\mathbf M}$.
This barrier region is the same 2DEG as in Fig.~\ref{fig:setup}(a) but modified by a magnetic proximity effect and a potential change. 
The change of the barrier Fermi contour with respect to the Fermi contour of the leads when the in-plane ${\mathbf M}$ is varied affects the transmission rates and produces ${\mathbf M}$-dependent changes in the device resistance. Furthermore, the barrier Fermi contour 
undergoes a deformation that increases with the strength of the Rashba SOC. Such a deformation allows for lead-barrier Fermi contour matching enabling multiple states to achieve high-transmission rates. 

The realization of perfect transmission can be intuitively understood by considering the action of the $\mathcal{P}_s\mathcal{T}\equiv 
\mathcal{P}\sigma_z\mathcal{T}$ operator on an incident wave with a given in-plane spin, as schematically shown in Fig.~\ref{fig:setup}(d). $\mathcal{T}$ reverses both the spin and motion of the incident wave, while 
$\mathcal{P}_s\equiv  
\mathcal{P}\sigma_z$ inverts both the spin (through the action of the Pauli matrix, $\sigma_z$) and position (through the action of space inversion $\mathcal{P}$) of the wave. As a result, by applying the $\mathcal{P}_s\mathcal{T}$ operator the incident wave on the left is transformed to itself, but as a transmitted wave on the right. Therefore, scattering states which are eigenfunctions of $\mathcal{P}_s\mathcal{T}$ experience perfect transmission. 

While our analysis is mostly focused on the proximity-induced magnetization in a 2DEG, the approach we consider can be also applied to other systems. In this work this is illustrated 
on the example of topological insulators in which topological surface states could be accompanied by trivial bands characteristic for Rashba SOC.

Following this introduction, in Sec.~II we describe the Hamiltonian and scattering states for considered heterostructures.
In Sec.~III we provide the expressions for conductance and transmission as well as their helicity-resolved components.   
The transmission resonances are connected to the spin-parity-time symmetry of the scattering states. The Fermi contour analysis 
in Sec.~IV reveals the role of wave vector and spin mismatch on the evolution of conductance and transmission with the direction 
and magnitude of the proximity-induced magnetization in a 2DEG. In Sec.~V we analyze the angular dependence and resonant behavior 
of the TAMR. In Sec.~VI we discuss how TAMR signatures could be used to distinguish between the topological and trivial states in heterostructures with topological insulators as well as note some open questions for future work.

\section{II. HAMILTONIAN AND SCATTERING STATES}

\begin{figure}[h]
\centering
\includegraphics*[trim=0cm 0cm 0cm 0cm,clip,width=6.56cm]{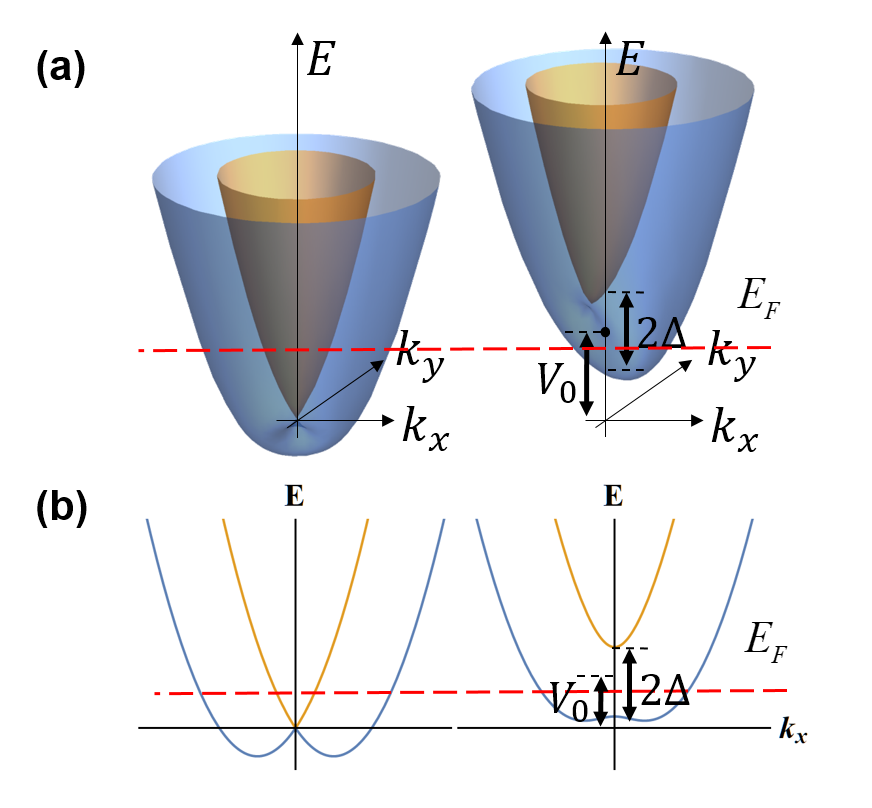}
\caption{(a) Typical band structure for the states in the lead (left) and barrier (right), see Fig.~\ref{fig:setup}. 
(b) The same band structure at normal incidence $k_y=0$. Red line: the Fermi energy.}
\label{fig:band structure}
\end{figure}

The model Hamiltonian of the system represented in Fig.~\ref{fig:setup}(a) is given by
\begin{equation}
H = \frac{p^2}{2m^\ast}+ \frac{\alpha}{\hbar}\left(\boldsymbol{\sigma}\times\mathbf{p}\right)\cdot\mathbf{\hat{z}}
+ [V_0 - \Delta(\mathbf{m} \cdot \boldsymbol{\sigma})]h(x),
\label{eq:Hamiltonian}
\end{equation}
where  $m^\ast$ is the effective mass, $\alpha$ is the Rashba SOC strength, $\mathbf{\hat{z}}$ is the unit vector along the $z$ axis, $\mathbf{p}=(p_x,p_y)$ is the 2D momentum operator, $\boldsymbol{\sigma}$ is the vector of Pauli matrices, $V_0$ describes the potential barrier, and $\Delta$ and $\mathbf{m}$ are the magnitude and direction of the proximity-induced ferromagnetic exchange field. 
The function $h(x)=\Theta(d/2+x)\Theta(d/2-x)$ describes a square barrier of thickness $d$. 
We focus on electrons, not holes~\cite{Winkler:2003,Rozhansky2016:PRB,Liu2018:PRL,Arovas1998:PRB}, with the typical
band structure shown in Fig.~\ref{fig:band structure}.

Due to  Rashba SOC, the wave functions can be classified by the helicity index, 
where $\lambda=1~(-1)$ refers to the inner (outer) Fermi contour as depicted in Fig.~\ref{fig:setup}(c).
The scattering states for the finite square barrier model can be written as ${\psi _\lambda }\left( {x,y} \right) = (1/\sqrt {2S}){e^{i{k_y}y}}{\phi _\lambda }\left( x \right)$, 
with sample area $S$ and the conserved parallel component of the wave vector $k_y$, 

\begin{widetext}
\begin{equation}
	{\phi _\lambda }\left( x \right) = \left\{ \begin{array}{ll}
	{e^{i{k_{x\lambda }}x}}\chi _\lambda ^{\left(  +  \right)} + {r_{\lambda \lambda }}{e^{ - i{k_{x\lambda }}x}}\chi _\lambda ^{\left(  -  \right)} + {r_{\lambda \bar \lambda  }}{e^{ - i{k_{x\bar \lambda  }}x}}\chi _{\bar \lambda  }^{\left(  -  \right)}, & x < -d/2, \hfill \\
	\sum\limits_{\lambda ' =  \pm 1} {\sum\limits_n {A_{\lambda \lambda '}^{\left( n \right)}{e^{i\widetilde k_{x\lambda '}^{(n)}x}}\tilde \chi _{\lambda '}^{\left( n \right)}} } , & -d/2 < x < d/2, \hfill \\
	{t_{\lambda \lambda }}{e^{i{k_{x\lambda }}x}}\chi _\lambda ^{\left(  +  \right)} + {t_{\lambda \bar \lambda  }}{e^{i{k_{x\bar \lambda  }}x}}\chi _{\bar \lambda  }^{\left(  +  \right)}, & x > d/2. \hfill \\ 
	\end{array}\right.
	\label{eq:States_finite_barrier}
	\end{equation}
\end{widetext}

Here the states in the lead are the same as those in the $\delta$-barrier model discussed in Appendix~A. In the barrier, the spinors are defined as
\begin{equation}
\widetilde \chi _{\lambda'} ^{\left( n \right)}{\text{ = }}\left( {\begin{array}{*{20}{c}}
	1 \\ 
	{ - i{\lambda'} {e^{i\widetilde \theta _{\lambda'} ^{\left( n \right)}}}} 
	\end{array}} \right),
\label{eq:barrier_spinor}
\end{equation}
where $\widetilde \theta _{\lambda'} ^{\left( n \right)}{\text{ =  arctan}}\left( {\frac{{\widetilde k_{x{\lambda'} }^{\left( n \right)} + {\Delta _y}/\alpha }}{{{k_y} - {\Delta _x}/\alpha }}} \right)$ and ${\widetilde k_{x{\lambda'} }^{\left( n \right)}}$ is the nth root ($n=1,\cdots,4$) of
\begin{equation}
\begin{gathered}
\frac{{{\hbar ^2}}}{{2m^\ast}}\left( {\widetilde k_{x{\lambda'} }^2 + k_y^2} \right) + {V_0}  \quad\quad\quad\quad\quad\quad\quad\quad\quad\quad\quad\\ 
+ {\lambda'} \sqrt {{{\left( {\alpha {{\widetilde k}_{x{\lambda'} }} + {\Delta _y}} \right)}^2} + {{\left( {\alpha {k_y} - {\Delta _x}} \right)}^2}} = E. \\ 
\end{gathered}
\label{eq:barrier_dispersion}
\end{equation}

For the  wave function in the barrier, all possible eigenstates should be involved. In Eq.~(\ref{eq:barrier_dispersion}), the total number of solutions for both $\lambda' = \pm1$ is four, but there are two possible situations for the distribution of the roots: All four roots belong to the $\lambda'= -1$ branch and no root to $\lambda'= 1$, or two roots belong to the $\lambda'=-1$ branch while another two belong to the $\lambda' = 1$ branch. 

At the boundaries ($x = -d/2$ and $x = d/2$), the wave function and its first derivative 
should be continuous, which leads to a group of linear equations.
By solving these linear equations, one can obtain the transmission coefficients $t_{\lambda \lambda}$ and  $t_{\lambda \bar  \lambda}$
for the square barrier model.

\section{III. Conductance and Transmission}
\label{Sec:Tunneling Conductance}

From the continuity equation, one can derive the particle current density
\begin{equation}
\bm{j} = \operatorname{Re} \left[{\psi ^\dag }\left( {\frac{\bm{p}}{m^\ast} + \frac{\alpha }{\hbar }\left( {{\bm{e}_z} \times \bm{\sigma} } \right)} \right)\psi \right].
\label{eq:current1}
\end{equation}

Inserting the scattering state in the right lead at $x>d/2$ from Eq.~(\ref{eq:States_finite_barrier}), 
we can obtain the particle current density of the $\lambda$ channel
\begin{equation}
{j_\lambda }{\text{ = }}\operatorname{Re} \left[ v/S{\left( {{{\left| {{t_{\lambda \lambda }}} \right|}^2}\cos {\theta _\lambda } + {{\left| {{t_{\lambda \bar \lambda }}} \right|}^2}\cos {\theta _{\bar \lambda }}} \right)} \right];
\label{eq:current2}
\end{equation}
here, the group velocity of the scattered particle, $v = \sqrt {( {\alpha/\hbar)^2 + (2E/m^\ast)}}$, has the same magnitude for two bands. 
This current contains contributions from
the intra- and interchannel transmission, where  $\theta_\lambda$ is the incident and $\theta_{\bar{\lambda}}$ the propagation angle of the cross-channel wave with the conservation of the $k_y$ component and 
$\theta _{\bar{\lambda}}$ is related to $\theta _{ \lambda }$ by $\cos {\theta _{\bar{\lambda}}} = {\sqrt {k_{\bar{\lambda}}^2 - k_\lambda ^2{{\sin }^2}\theta_{\lambda} } }/{k_{ \bar{\lambda}}}$.

The longitudinal tunneling charge current of the $\lambda$ channel can be written as 
\begin{equation}
{I_\lambda } = eD\sum\limits_{{k_x} \geqslant 0,{k_y}} {{j_\lambda }\left[ {f\left( {E - eV} \right) - f\left( E \right)} \right]},
\label{eq:charge_current}
\end{equation}
where $D$ is the width of the sample and $f(E)$ is the Fermi-Dirac distribution. In the low-bias limit, i.e., $\left| {eV} \right| \ll {E_F}$, one can use  
the approximation $f\left( {E - eV} \right) - f\left( E \right) \approx eV\delta \left( {E - E_F} \right)$.

Performing the following replacement
\begin{equation}
\sum\limits_{{k_x} \geqslant 0,{k_y}} {}  \leftrightarrow \frac{S}{{{\left( {2\pi } \right)}^2}}\int_0^\infty  {d{k_x}} \int_{ - \infty }^\infty  {d{k_y}},
\label{eq:sum_integral}
\end{equation}
and inserting the particle current density in Eq.~\eqref{eq:current2}, we get the expression for the conductance $G_\lambda  \equiv I_\lambda /V$ in the $\lambda$ channel 
\begin{equation}
{G_\lambda } = \frac{e^2}{h}\frac{D}{2\pi}\int_{ - \frac{\pi }{2}}^{\frac{\pi }{2}} {d{\theta _\lambda }k_F^\lambda {T_\lambda }\cos {\theta _\lambda }},
\label{eq:conductance1}
\end{equation}
where $k_F^\lambda$ is the Fermi wave vector of the $\lambda$ channel and
the transmission is 
\begin{equation}
{T_\lambda } = \operatorname{Re} \left[{\left| {t_{\lambda \lambda}} \right|^2} + {\left| {t_{\lambda \bar \lambda }} \right|^2}\left( {\cos {\theta _{\bar \lambda }}/\cos {\theta _\lambda }} \right)\right].
\label{eq: transmission}
\end{equation}
The total conductance is the sum of the two channels, 
\begin{equation}
{G} = \sum\limits_{\lambda  =  \pm 1} {G_\lambda }.
\label{eq:total_conductance}
\end{equation}

As shown in Fig.~\ref{fig:setup}, the $\mathcal{P}_{s}\mathcal{T}$ symmetry leads to perfect transmission. Therefore, transmission resonances occur whenever the scattering states are such that $\mathcal{P}_{s}\mathcal{T}\psi(x,y)=\xi\psi(x,y)$, with eigenvalues of the form $\xi=e^{i\eta}$. However, $\mathcal{P}_{s}\mathcal{T}$ does not commute with the Hamiltonian in Eq.~(\ref{eq:Hamiltonian}). Therefore, it is not an intrinsic symmetry of the system. Instead, the $\mathcal{P}_{s}\mathcal{T}$ symmetry emerges only for certain specific system parameters and scattering states satisfying,
\begin{equation}
[H,\mathcal{P}_{s}\mathcal{T}]\psi_R(x,y)=0,
\label{eq:commutator}
\end{equation}
where the index $R$ emphasizes that the relation holds only at resonances.
This symmetry generalizes a simple case of resonances in a potential barrier (or a spinless) system~\cite{Griffiths:2005}. 
We will further analyze this occurrence of resonances in Appendix~B.

\section{IV. FERMI CONTOUR ANALYSIS}
\subsection{A. Wave vector and spin mismatch}

\begin{figure}[b]
\centering
\includegraphics*[trim=0.6cm 0.6cm 0.6cm 0.4cm,clip,width=8.5cm]{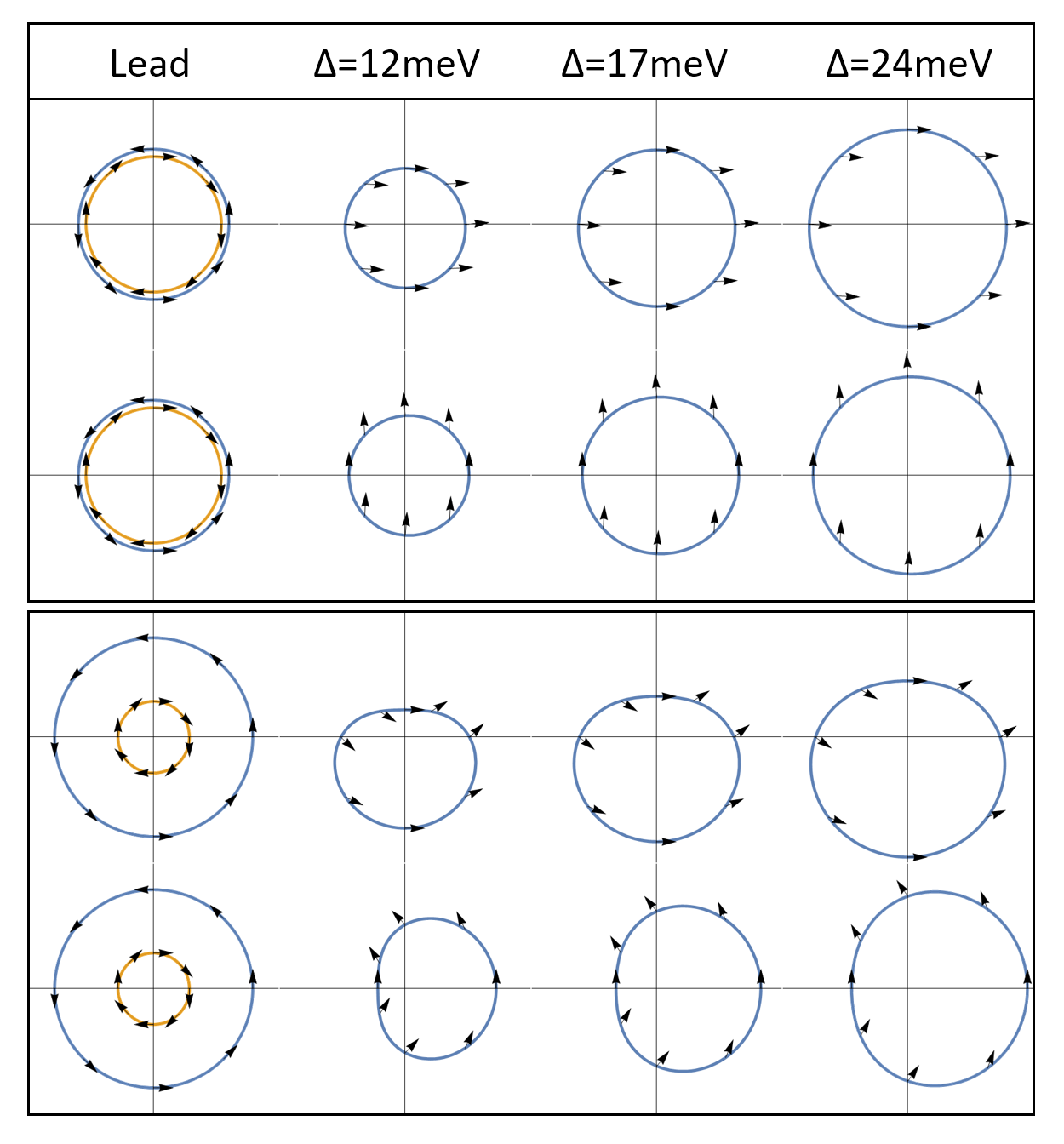}
\caption{Fermi contours in the lead (leftmost) 
and  in the barrier when $V_0=15$~meV and $\Delta=12$, 17, and 24~meV. The upper (lower) panel is for 
$\alpha=0.093$~eV\AA~($\alpha=0.93$~eV\AA). In each panel, the first row 
corresponds to $\mathbf{m}\parallel \mathbf{x}$ and the second row  
$\mathbf{m}\parallel \mathbf{y}$. The blue and yellow contours denote lower and upper bands, respectively, and inside the barrier the 
upper band disappears since its bottom is above the Fermi energy $E_F$. The spin orientations are represented by arrows.}
\label{fig:fermi contours}
\end{figure}

To understand the relation between conductance and barrier parameters, it is convenient to perform a Fermi contour analysis. 
The basic idea is that the better the matching between Fermi contours in the lead and barrier, the higher the transmission.  
We recall that for a simple $\delta$ barrier, the transmission 
is $T{\text{ = 1/}}\left( {1 + {Z^2}} \right)$, where $Z$ 
is the effective barrier strength which combines the influence of a native barrier and the mismatch of the 
Fermi wave vectors in the two regions~\cite{Zutic1999:PRBa,Vezin2020:PRB}.
Since the difference between the lead and barrier Fermi contours is associated with the effective 
barrier potential, a larger mismatch between Fermi contours corresponds to a larger effective potential $Z$ and thus to a low transmission.

In the presence of Rashba SOC the effective interfacial barrier is inequivalent for two helicities (for outer/inner Fermi contours) leading to an
important influence of spin mismatch on transmission~\cite{Vezin2020:PRB}.  
For the spin mismatch, we can decompose the incident spinor 
in 
the basis constructed by the corresponding barrier eigenspinor $\chi_\uparrow$ and its antiparallel partner $\chi_\downarrow$, i.e., $\left | \chi_\text{in} \right\rangle  =  \left\langle \chi_\uparrow  {\left |\right.}\chi_\text{in} \right\rangle 
\left | \chi _\uparrow \right\rangle+\left\langle \chi _\downarrow  {\left |\right.} \chi _\text{in} \right\rangle \left | \chi _\downarrow \right\rangle$.
The first term undergoes a weak effective barrier while the second term experiences a strong barrier ~\cite{Vezin2020:PRB}. Therefore, considering the spin mismatch, we need to include a correction of ${\left| {\left\langle {{\chi _ \uparrow }|{\chi _{{\text{in}}}}} \right\rangle } \right|^2}$ in the transmission, i.e.,
$T \approx {\left| {\left\langle {{\chi _ \uparrow }|{\chi _{{\text{in}}}}} \right\rangle } \right|^2} T_{\chi_\uparrow}$, where 
$T_{\chi_\uparrow}$ is the transmission
when spin mismatch is ignored.
When the state inside the barrier is the same (up to a phase) as that in the lead, perfect transmission is achieved.

The band structure of 
the lead and barrier states shown in Fig.~\ref{fig:band structure} can be further
analyzed for parameters of a typical InGaAs/InAlAs 2DEG with $m^\ast = 0.05 m_0$, where
$m_0$ is the free-electron mass, and $\alpha = 0.093\;{\rm eV \AA}$~\cite{Nikolic‡2005:PRB}. The corresponding Fermi contours
with spin orientations for various strength of the proximity-induced exchange and a tenfold increase in SOC are shown in Fig.~\ref{fig:fermi contours}.
While in the leads, the Fermi 
contours of the Rashba bands are perfectly circular,  inside of the barrier they are shifted and distorted in the 
direction $\perp$ to the magnetization $\mathbf{M}$, which is
along the direction of the proximity-induced exchange field $\mathbf{m}$ shown in Fig.~\ref{fig:fermi contours} for $\mathbf{m}\parallel \mathbf{x}$ and  $\mathbf{m}\parallel \mathbf{y}$. 
These effects can be seen more clearly in the large $\alpha$ states. The spin is polarized along the $\mathbf{M}$ in the barrier.

To understand the behavior of the conductance, the overall matching of the Fermi contours should be considered, i.e., not only the matching of individual states but also the overall matching in the radius and curvature of the contours. When the incoming particles are from two Fermi contours, we need to compare both the inner and outer lead contours with the barrier contour.

\subsection{B. Conductance peak near \boldmath$\Delta = V_0$}

\begin{figure}[t]
\centering
\includegraphics*[trim=0cm 0.4cm 0.2cm 1.4cm,clip,width=8.7cm]{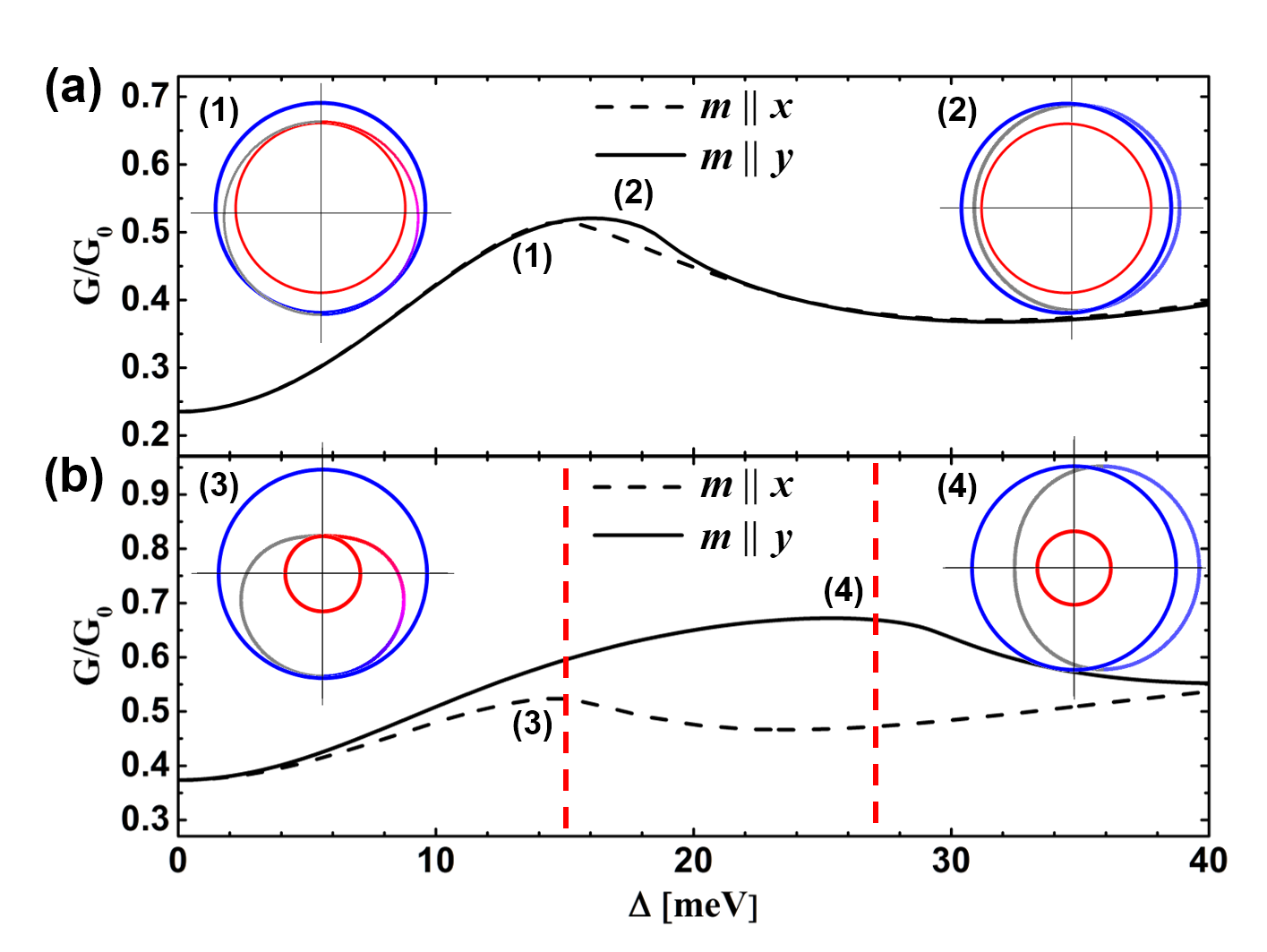}
\caption{(a) Dependence of the conductance on $\Delta$ with $E_F=10$~meV, $V_0=15$~meV, $d=13$~nm, and $\alpha=0.093$~eV\AA, where the dashed and solid curves denote the conductance for $\mathbf{m} \parallel \mathbf{x}$ and $\mathbf{m} \parallel \mathbf{y}$, respectively. 
Insets (1) and (2) 
show the matching between Fermi contours in the lead (blue and red) and in the barrier (multi-color) at labeled peaks. (b) The same as (a) but $\alpha$ is 10 times greater.
The dashed vertical lines labeled by (3) and (4) denote the values of $\Delta$ for which the conductance maxima are achieved and
the corresponding Fermi contours are given in the two insets. }
\label{fig:peaks}
\end{figure}

To explore the evolution of conductance as a function of proximity-induced field and its direction, throughout this work
we consider its normalized value in Eq.~(\ref{eq:total_conductance}) expressed in terms of the Sharvin conductance,
\begin{equation}
{G_0} = \frac{e^2}{h}\frac{2D}{\pi}\sqrt{{{\left( {\frac{\alpha \, m^\ast}{\hbar ^2}} \right)}^2} + \frac{2m^\ast \,E_F}{\hbar ^2}}.
\label{eq:Sharvin}
\end{equation}
Similar to the examined changes in the Fermi contours from Fig.~\ref{fig:fermi contours},
we focus in Fig.~\ref{fig:peaks}(a) on the conductance for $\mathbf{M}$ transverse ($\mathbf{m}\parallel \mathbf{x}$) 
and parallel ($\mathbf{m}\parallel \mathbf{y}$) to the barrier [recall Fig.~\ref{fig:setup}(d)], $G(\phi=0)$ and $G(\phi=\pi/2)$, respectively. 
The conductance exhibits a nonmonotonic behavior with maxima, labeled by (1) and (2), occurring at different $\Delta$ and accompanied by 
the corresponding Fermi contours in the insets (1) and (2). 
To understand this behavior it is useful to consider a small SOC limit, 
$\alpha \ll \Delta/\sqrt {(2m^\ast/\hbar ^2)\left( {E_F + \Delta  - V_0} \right)} $,
where the barrier Fermi contour can be approximately written as
\begin{equation}
	\begin{gathered}
		{\left( {k_x - \frac{{m^\ast \alpha }}{{\hbar ^2}}\sin \phi } \right)^2} + {\left( {k_y + \frac{m^\ast \alpha}{\hbar ^2}\cos \phi } \right)^2} \hfill \\
		\hspace*{90pt} \approx \frac{2m^\ast}{\hbar ^2}\left( {E_F + \Delta  - V_0} \right). \hfill \\
	\end{gathered} 
	\label{eq:barrier_contour}
\end{equation} 
This implies that the barrier contour for small $\alpha$ reduces to a shifted circle with radius $r_0\sqrt{\left( {{E_F} + \Delta  - {V_0}} \right)/{E_F}}$, 
where $r_0$ is the average radius of the inner and outer lead circles. In the region near the line $V_0=\Delta$, both $G(0)$ and $G(\pi/2)$ reach their maxima 
because the best matching of Fermi contours between lead and barrier is obtained. Specifically, when $\alpha  \to 0$, $V_0=\Delta$ leads to a perfect transmission. We 
notice that the shift  of barrier circle, which is always $\perp$ to $\mathbf{m}$,  is of the first order in $\alpha$, while the deformation is, at least, a second order correction.

A modest shift between the two conductance maxima in Fig.~\ref{fig:peaks}(a) is largely enhanced in Fig.~\ref{fig:peaks}(b), when the SOC 
is increased tenfold and $\alpha = 0.93\;{\rm eV \AA}$. As shown in Figs.~\ref{fig:peaks}(a) and ~\ref{fig:peaks}(b), 
$G(0)$ is peaked exactly at $V_0=\Delta$, where the contours from insets (1) and (3) 
share the least overall mismatch. In the small $\alpha$ limit, the barrier contour is shifted downward and its radius 
equals to the average radius of the inner and outer lead contours. 
Due to the asymmetry of the barrier Fermi contour, the best match for both the upper and lower half of the circle cannot be achieved simultaneously. 
Therefore, the upper (lower) half of the barrier contour tends to match the inner (outer) lead contour since the spin mismatch is smaller for these states.
Thus the simultaneous achievement of these large transmission conditions 
at $V_0=\Delta$ results in the maximum conductance. This statement still holds for large $\alpha$, so generally $V_0=\Delta$ is always the maximum condition for $G(0)$.

However, the maximum of $G(\pi/2)$ is achieved when the barrier circle shares the same size as the outer 
circle of the lead from Fig.~\ref{fig:peaks} in insets (2) and (4). 
Up to the first order in $\alpha$, the condition leading to the maximum $G(\pi/2)$ is given by $\Delta  = {V_0} + \alpha k_F$. When such a condition is satisfied, the barrier Fermi contour matches the outer lead contour instead of the inner one, because the main contribution of the conductance is from the incident particles on the outer Fermi contour. 

\begin{figure}[t]
\vspace{-0.3cm}
\centering
\includegraphics*[width=9cm]{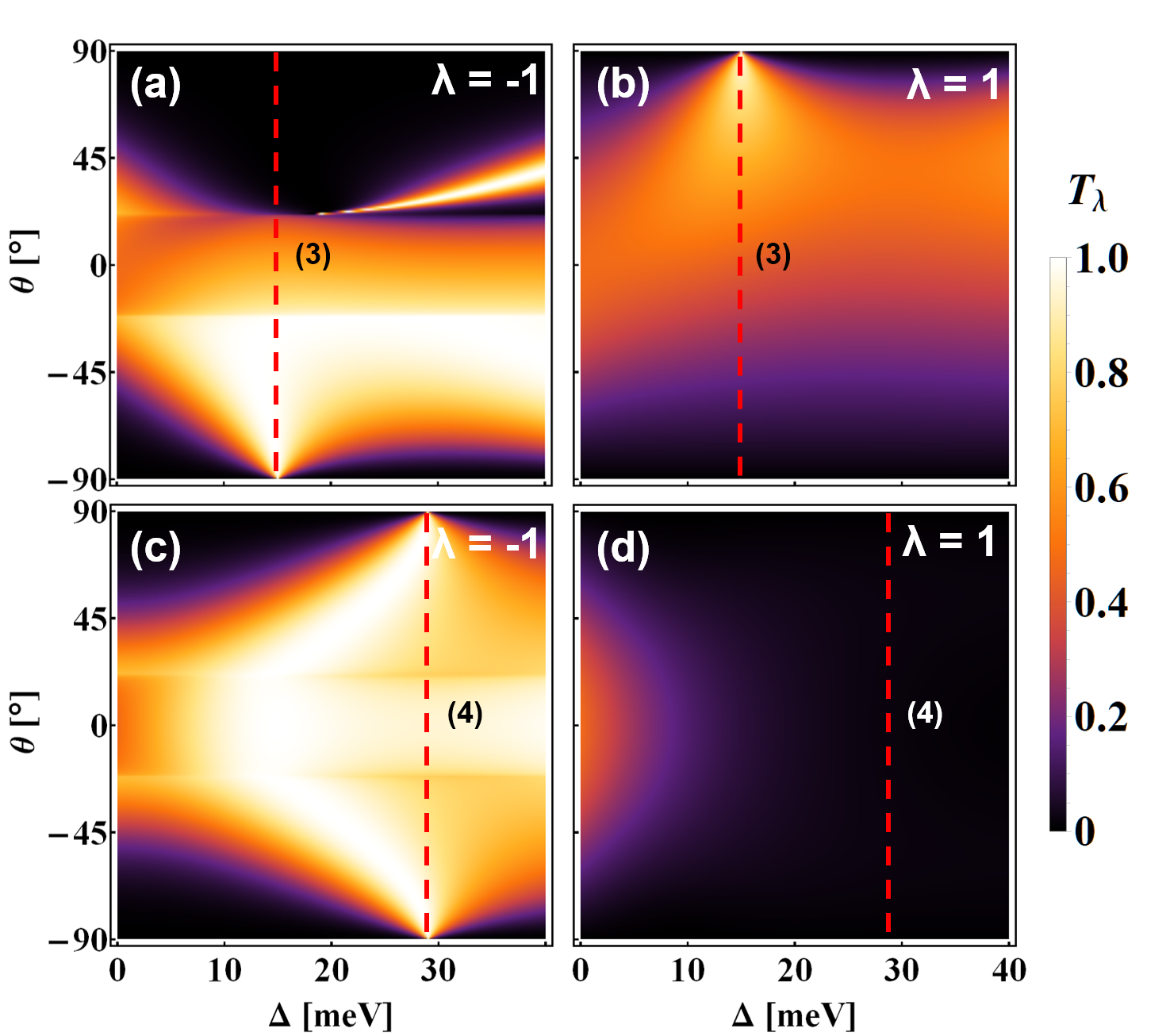}
\caption{Transmission $T_\lambda$ from each energy band (a), (b) when $\mathbf{m} \parallel \mathbf{x}$ 
and (c), (d)  when $\mathbf{m} \parallel \mathbf{y}$, as a function of $\Delta$ and incident 
angle $\theta$, with $E_F=10$~meV, $V_0=15$~meV, $d=13$~nm, and $\alpha=0.93$~eV\AA, as in Fig.~\ref{fig:peaks}(b). 
The dashed vertical lines labeled by (3) and (4) denote the values of $\Delta$ for conductance maxima from Fig.~\ref{fig:peaks}(b).
}
\label{fig:resonance}
\end{figure}

The previously discussed nonmonotonic behavior of the conductance is a consequence of the collective contributions of multiple resonant states 
corresponding to different propagation directions of the tunneling carriers.  
For a strong SOC in Fig.~\ref{fig:peaks}(b), the appearance of a large number of resonant states at the value of $\Delta$ indicated by vertical  
lines results in the maximum of conductance labeled by (3) and (4). With the change of the $\mathbf{m}$ the parameter space where resonances 
emerge is shifted, leading to a shift in the corresponding $\Delta$ value. 

The origin of the conductance maximum at (3) and (4) 
can be understood from Fig.~\ref{fig:resonance} by examining  the transmission at $E_F$,
$T_\lambda$, which reveals multiple resonances, depending on $\Delta$ and the incident angle $\theta$.
As shown in Fig.~\ref{fig:resonance}, where the corresponding values $\Delta$ for the conductance
maxima (3) and (4) from Fig.~\ref{fig:peaks}(b) are given again,
the dominant contribution to $T_\lambda$ comes
from $\lambda=-1$ channel, 
which corresponds to the transport via the outer Fermi contour states in the leads. 
The region of bright colors in Figs.~\ref{fig:resonance}(a) and ~\ref{fig:resonance}(c) shows strong  $T_{-1}$ which approximately satisfies the 
symmetry condition Eq.~(\ref{eq:commutator}), further analyzed in Appendix~B and connected to the occurrence 
of different resonances. 

The observed behavior in 
Figs.~\ref{fig:peaks}(b), \ref{fig:resonance}(a), and \ref{fig:resonance}(c) leads us to conclude that the maximum of the total conductance is determined mainly by 
the range of $\theta$ for dominant transmission. For $G(0)$ or $\mathbf{m}\parallel \mathbf{x}$, the Fermi contour inside the barrier shifts vertically and the condition 
for maximum transmission angle is achieved exactly at $\Delta=V_0$, when the shift of the energy band by the potential $V_0$ inside the barrier 
is canceled by the Zeeman shift $-\Delta$ for the $\lambda=-1$ channel.

\begin{figure}[t]
\centering
\includegraphics*[trim=0.4cm 0.2cm 0.9cm 0.4cm,clip,width=8.7cm]{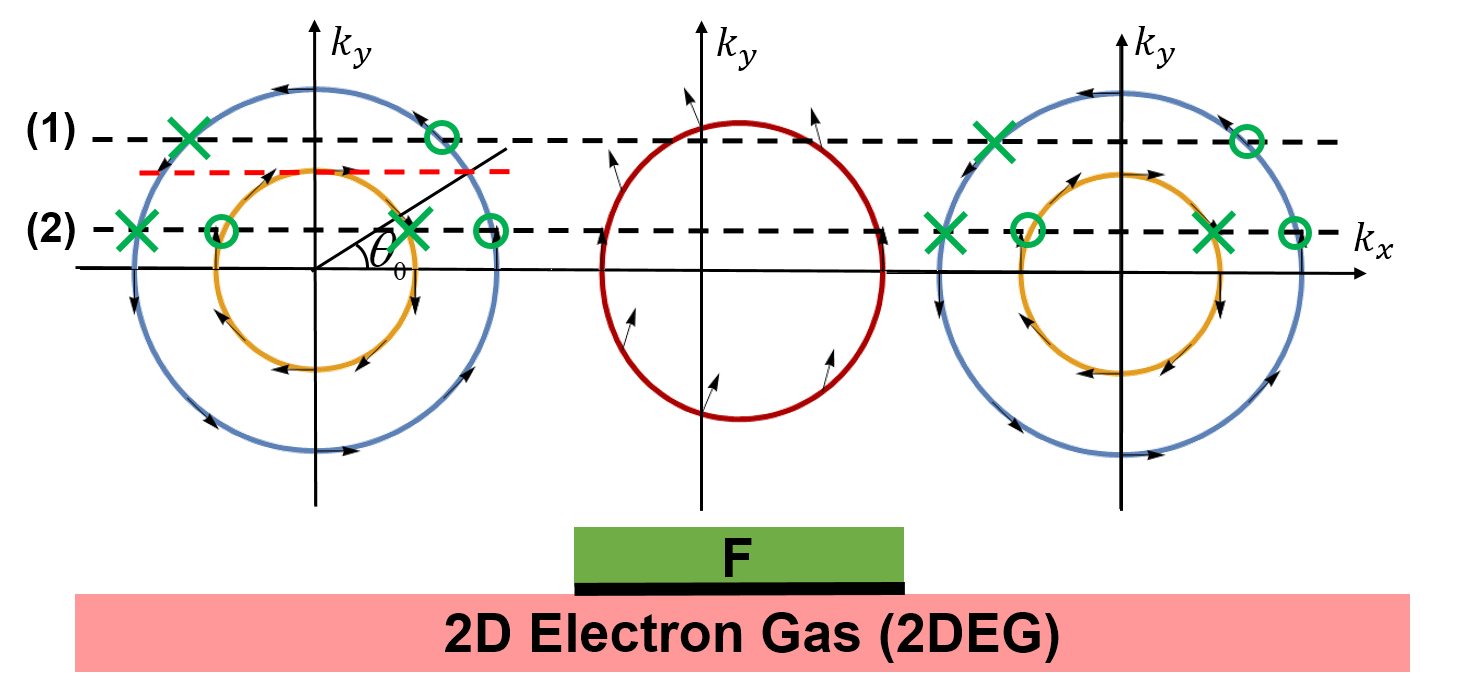} 
\caption{Planar geometry and schematic 
diagram of scattering processes. 
The current flows only through a 2DEG which is modified in the barrier region by the electrostatic potential and the proximity-induced spin splitting 
from the ferromagnet, F;
Fermi contours correspond to the states in the left lead, barrier, and right lead. Dots (crosses) denote the states favored (unfavored) in the scattering. 
States with positive $k_x$ are transmitted, while those with negative $k_x$ are reflected.}
\label{fig:back}
\end{figure}

In contrast, for $G(\pi/2)$ or $\mathbf{m}\parallel \mathbf{y}$,  
the best contour matching at large angles is achieved for an enlarged Fermi contour inside the barrier at $\Delta\approx V_0+\alpha k_F$
as shown in inset (4) of Fig.~\ref{fig:peaks}(b) and discussed before. 
This is consistent with Fig.~\ref{fig:resonance}(c) revealing an almost perfect transmission near the grazing incidence ($\theta \approx \pm 90^\mathrm{o}$),
rather than just a maximum $T_{-1}$ expected to occur near the normal incidence ($\theta \approx 0^\mathrm{o}$).  
We can understand this behavior of $T_{-1}$ and $G(\pi/2)$ from the allowed scattering processes illustrated in Fig.~\ref{fig:back}.
Considering the spin matching, reflection to the band with opposite helicity is favored, while the situation for transmission is the opposite. On the other hand, for carriers from 
the $\lambda=-1$ band (outer contour), when the incoming angles are greater than the critical angle ${\theta _0} =  \pm \arcsin \left( k_1/{k_{ - 1}} \right)$, the transmission and reflection 
to the $\lambda=1$ band are not allowed because there are no such propagating states as shown for dashed line (1). Therefore, in this regime, back scattering is suppressed while $T_{-1} $ 
is enhanced. To ensure that states with such large incident angles are transmitted across the barrier, 
the corresponding barrier eigenstates should be propagating, implying that the contour should at least have the same size as the outer lead contour. 
Since the matching deteriorates when the size of the barrier contour increases, the maximum of $G(\pi/2)$ is obtained when the barrier contour is the same size as the outer lead contour, 
confirming again $\Delta \approx V_0 + \alpha k_F$.

\section{V. Tunneling Anisotropic Magnetoresistance (TAMR)}

\subsection{A. Infinite vs finite geometry}

This difference for the maximum of the total conductance as a function of $\mathbf{m}$ is the origin of the strong TAMR.
To characterize the strength of the anisotropic response, we introduce the in-plane 
TAMR coefficient\cite{Moser2007:PRL,MatosAbiague2009:PRBa},
\begin{equation}
{\rm TAMR}(\phi)=\frac{R(\phi)-R(0)}{R(0)}=\frac{G(0)-G(\phi)}{G(\phi)},
\label{eq:tamr-def}
\end{equation}
with $\phi$ defined in Fig.~1(d). Unlike the MR values, 
which may change considerably from sample to sample, the TAMR coefficient is known  
to be more robust against specific sample details~\cite{DW}. Up to the second order in the SOC strength, the extreme values of
the ${\rm TAMR}(\phi)$ are given by the contrast between the MR measured for 
{\bf m} parallel and perpendicular to the current, i.e., ${\rm TAMR}(\phi = \pi/2)$. For brevity, we use 
\begin{equation}
{\rm TAMR} \equiv {\rm TAMR}(\phi = \pi/2), 
\label{eq:tamr}
\end{equation}
unless $\phi$ is explicitly specified. 

Our previous analysis, based on the model from Sec.~II, relies on an 
infinite heterostructure and its translational invariance perpendicular to the current.
It would then be important to examine if the observed trends are retained by relaxing this assumption 
within a finite geometry. 
To consider this, our numerical calculations 
are based on a finite-difference scheme leading to the discretization of the Hamiltonian in Eq.~(\ref{eq:Hamiltonian}), which can then be written in the tight-binding representation as,
\begin{widetext}
\begin{eqnarray}
\label{eq:tb-H}
H_{\rm TB}&=&\sum_{\mathbf{n},\sigma} \left(4t-V_{0}h_\mathbf{n}\right)|\mathbf{n},\sigma\rangle\langle \mathbf{n},\sigma| - t\sum_{\langle \mathbf{n},\mathbf{n}'\rangle,\sigma}|\mathbf{n},\sigma\rangle\langle \mathbf{n}',\sigma|-\frac{2 \Delta}{\hbar}\sum_{\mathbf{n},\sigma,\sigma'}\left(\mathbf{m}\cdot \mathbf{S}_{\sigma\sigma'}\right)h_{\mathbf{n}}|\mathbf{n},\sigma\rangle\langle \mathbf{n},\sigma'|	\\ \nonumber
	&+&\frac{\alpha}{\hbar a}\sum_{\mathbf{n},\sigma,\sigma'}\left(i\; S_{\sigma\sigma'}^{x}|\mathbf{n},\sigma\rangle\langle \mathbf{n}+\mathbf{e}_y,\sigma'|-i\; S_{\sigma\sigma'}^{y}|\mathbf{n},\sigma\rangle\langle \mathbf{n}+\mathbf{e}_x,\sigma'|+ {\rm H.c.}\right) 	 
\end{eqnarray}	
\end{widetext}
where $\langle \mathbf{r}|\mathbf{n},\sigma\rangle = \psi_\sigma(a \mathbf{n})$ represents the wave function 
at sites $\mathbf{n}=(n_x,n_y)$ of the square lattice with lattice constant $a$, $t=\hbar^2/2m^\ast a^2$ is the hopping parameter, $\mathbf{e}_x$ ($\mathbf{e}_y$) is the unit vector along the $\mathbf{x}$ ($\mathbf{y}$) axis, and $\mathbf{S}_{\sigma\sigma'}$ is the vector of spin matrix elements, $S_{\sigma\sigma'}^{x,y}=\langle\sigma|\hbar \sigma_{x,y}/2|\sigma'\rangle$. In the second term on the right-hand side of Eq.~(\ref{eq:tb-H}), $\langle \mathbf{n},\mathbf{n'}\rangle$ indicates the sum is over nearest neighbors. The form of the barrier is determined by $h_{\mathbf{n}}=h(a \mathbf{n})$, which we assume to have a rectangular shape and width $d$. 

To compute the conductance, two semi-infinite metallic leads without Rashba SOC are attached to the scattering region described by Eq.~(\ref{eq:tb-H}). The numerical calculations were performed with the Kwant package~\cite{Groth2014:NJOP} for quantum transport, which allows for an efficient computation of the zero-temperature differential conductance by using the Landauer formula,
\begin{equation}
G=\frac{e^2}{h}\sum_{i\sigma\in l,j\sigma'\in r}|\mathcal{S}_{i\sigma,j\sigma'}|^2.
\label{eq:tb-G}
\end{equation}
Here $\mathcal{S}_{i\sigma,j\sigma'}$ represents the scattering matrix elements 
and the summation is over the conducting channels in the left ($l$) and right ($r$) leads.

\begin{figure}[h]
\centering
\includegraphics*[width=8.7cm]{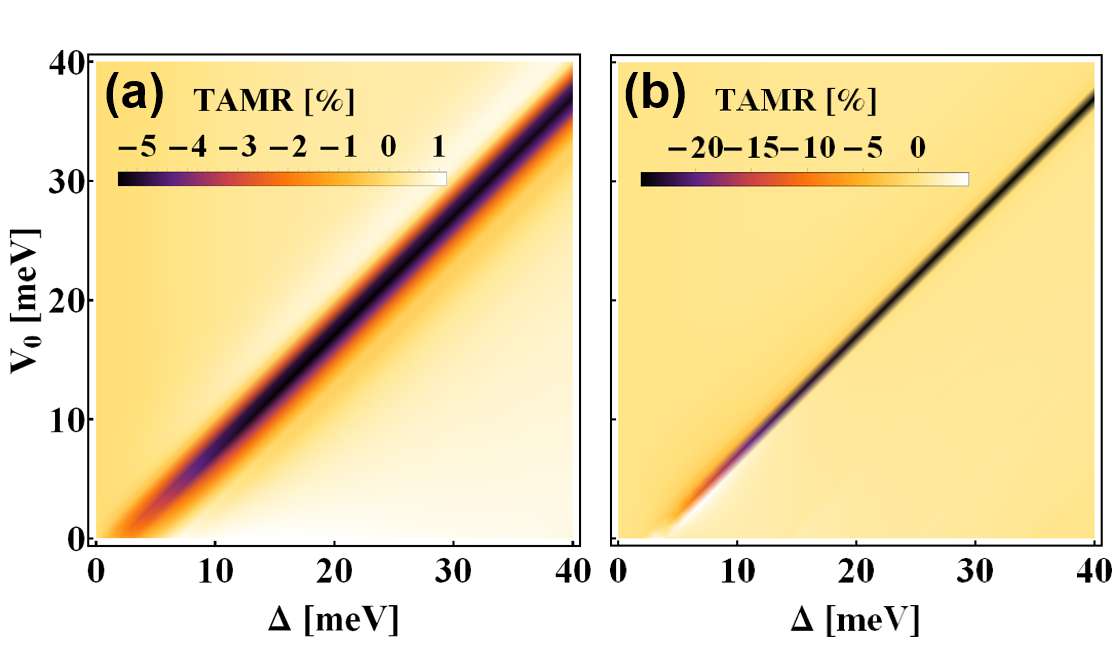}
\caption{(a) Dependence of TAMR amplitude on $V_0$ and $\Delta$ for 2DEG system with $d$=13~nm thick barrier, $E_F$=10~meV, and $\alpha=0.093$~eV\AA. 
(b) TAMR for a finite system with the same $d$ and $\alpha$, 
but with the hopping parameter $t = 0.112\; {\rm eV}$, corresponding to a lattice spacing 
of $a = 2.6\; {\rm nm}$, and the scattering region was discretized into a $100\times 40$ lattice. 
}
\label{fig:TAMR_VO_Delta}
\end{figure}

To examine the influence of geometry on the evolution of TAMR with $\Delta$ and $V_0$, we use the previous InGaAs/InAlAs 2DEG-based parameters 
and compare in Fig.~\ref{fig:TAMR_VO_Delta}(a) the results for an infinite geometry with those in Fig.~\ref{fig:TAMR_VO_Delta}(b) for a finite geometry.
In Fig.~\ref{fig:TAMR_VO_Delta}(a), the TAMR exhibits a sharply peaked behavior for $\alpha=0.093$~eV\AA $\;$  with extreme values along a line in the 
vicinity of $V_0=\Delta$. For a given value of $V_0$, the width of the TAMR peak is estimated as the difference between the values of $\Delta$ corresponding 
to the maximum of $G(0)$ and $G(\pi/2)$ which, according to our discussion in Sec.~IVB and Fig.~\ref{fig:peaks}, in the weak SOC limit is given by $\alpha k_F$. 
Complementary finite-size calculations in Fig.~\ref{fig:TAMR_VO_Delta}(b) have been performed calculations using Kwant~\cite{Groth2014:NJOP}  with TAMR 
obtained from Eqs.~(\ref{eq:tb-H}) and (\ref{eq:tb-G}). A finite system with a scattering region was discretized into a $100\times 40$ lattice with a spacing of 
2.6 nm and the hopping parameter $t = 0.112\; {\rm eV}$. The results, for the same range of $V_0$ and $\Delta$ as in Fig.~\ref{fig:TAMR_VO_Delta}(a), 
reveal that while finite-size effects slightly increase and sharpen the TAMR, the overall qualitative behavior remains similar to the calculations for an infinite system. 
This confirms the TAMR robustness mentioned above and the suitability of the considered infinite geometry. 
	
\subsection{B. Angular dependence}

Since the helical spin textures in 2DEG systems, depicted in Fig.~\ref{fig:setup}(c), are also inherent to the surfaces states of 3D topological insulators (TIs), it is important 
to examine if there are any differences between their respective TAMR signatures, for example, in the angular dependence of TAMR. This distinction could be very 
important since in 3D TIs, depending on the Fermi energy, their transport properties  may be dominated by topological surface states with a Dirac-like dispersion, 
trivial Rashba-like states, or both~\cite{Culcer2020:2DM,Chang2015:PRB,Trang2016:PRB,Douli2020:NL}. 

While we focus on in-plane TAMR, we note that out-of-plane TAMR calculated  in vertical structures with TIs can become large. Within an effective Hamiltonian 
description  it can be enhanced from $\sim 1$ \% to $\sim 15$ \% with a direct F/TI contact which opens the gap in the TI's surface state~\cite{Mahfouzi2014:PRB} 
and can even reach $\sim 50$ \%,  predicted from first principles~\cite{Marmolejo-Tejada2017:NL}.

\begin{figure}[h]
\centering
\includegraphics*[trim=0.1cm 0cm 0cm 0cm,clip,width=9cm]{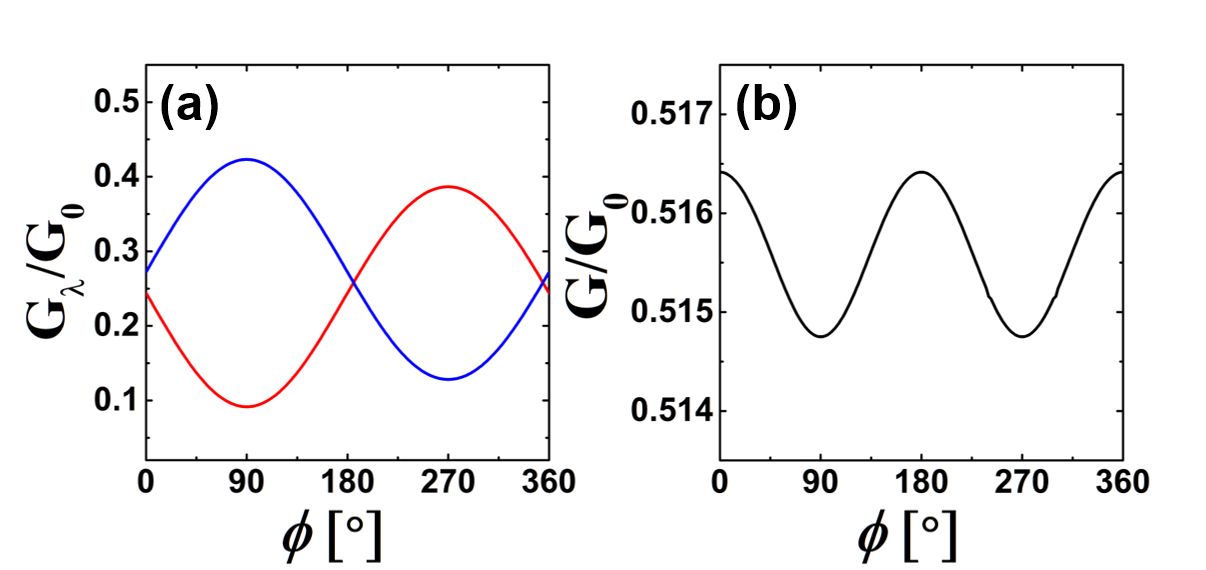}
\caption{ Angular dependence of  (a) channel-resolved conductance  (blue line: $\lambda=-1$  and red line: $\lambda=1$) 
and (b) total conductance for a 2DEG 
with the $\delta$ barrier, where $E_F=10$~meV, $\Delta=12$~meV, $V_0=15$~meV, $d=13$~nm, and $\alpha=0.093$~eV\AA.}
\label{fig:conductance_phi}
\end{figure}

To analyze in-plane angular dependence, it is helpful to use the $\delta$-barrier model 
from Appendix A and obtain some 
analytical results. For a 2DEG system,  in the limit 
of $\Delta/V_0 \ll 1$,  the leading contribution to the angular dependence of the conductance from the two incoming channels with helicity $\lambda=\pm1$,  
is  $\pm \sin \left( \phi  \right)$, as shown in Fig.~\ref{fig:conductance_phi}(a). However, with their opposite signs, these leading contributions cancel in the total 
conductance from Fig.~\ref{fig:conductance_phi}(b), which becomes significantly smaller, quadratic in the small parameter, and has a different angular dependence, 
resulting in 

\begin{equation}
\text{TAMR}_{\text{2DEG}} \left( \phi  \right) \sim (\Delta /V_0)^2 \sin^2 \left( \phi  \right).
\label{eq:TAMR_2DEG}
\end{equation}

A similar TAMR analysis can be performed for a 3D TI dominated by the Dirac-like topological surface state~\cite{Qi2011:RMP,Shen:2012},
where in the heterostructure from Fig.~\ref{fig:setup}(a) the 2DEG is replaced by a 3D TI. The corresponding spin-momentum locking 
of these surface states is used to explain various experiments~\cite{Qi2011:RMP,Shen:2012,Kandala:2019}, including unusual electrically tunable 
magnetoresistive effects in TIs with an applied magnetic field or magnetic doping~\cite{Sulaev2015:NL,Zhang2014:NC}.
Here we use the $\delta$-barrier model, based on the approach from Ref.~\cite{Scharf2016:PRL}, given by
\begin{equation}
H = {v_F}\left(\boldsymbol{\sigma}\times\mathbf{p}\right)\cdot\mathbf{\hat{z}} + [V_0 - \Delta(\mathbf{m} \cdot \boldsymbol{\sigma})]d \delta(x), 
\label{eq:TI_Hamiltonian}
\end{equation}
with $v_F$ the Fermi velocity, while the remaining quantities retain the meaning from Eq.~(\ref{eq:Hamiltonian}).
Since we consider an in-plane magnetization, the topological surface state remains gapless. 
We can then derive an approximate analytical TAMR expression for topological states which, up to the  second order in $\Delta/V_0$, yields 
\begin{widetext}
\begin{equation}
{\text{TAM}}{{\text{R}}_{{\text{TI}}}}\left( \phi  \right) \approx \frac{{\tan {Z_0}({Z_0} - 2\tan {Z_0})[\left( {3 - {{\cos }^2}{Z_0}} \right)\text{arctanh}(\cos{Z_0}) - 3\cos{Z_0}]}}{{2[\cos {Z_0} - {{\sin }^2}{Z_0}\text{arctanh}(\cos {Z_0})]}}{\left( {\frac{\Delta }{{{V_0}}}} \right)^2}{\sin ^2}\left( \phi  \right),
\label{eq:TAMR_TI}
\end{equation}
\end{widetext}
with $Z_0=V_0d/(\hbar v_F)$ the dimensionless barrier strength, used
also in studies of TIs with multiple F regions~\cite{Salehi2011:PE}.

While both Eqs.~(\ref{eq:TAMR_2DEG}) and (\ref{eq:TAMR_TI}) yield the same angular dependence, this could simply be a consequence of the assumed $\delta$-function barrier.
Instead we next consider a more realistic square barrier to check if the angular dependence in TAMR remains
the same for trivial and topological states.
For a 3D TI we choose ${{\left( {\text{Bi}_x}{\text{Sb}_{1 - x}} \right)}_2}\text{Te}_3$ 
with $x=0.36$, effective mass $m^\ast=0.27m_0$, 
Rashba SOC $\alpha=0.36$~eV\AA, Fermi velocity of the surface state $4\times10^5$ m/s, the energy difference between the Dirac point 
and the crossing point of the Rashba bands $\Delta E=250$ meV, and $E_F=260$ meV measured from the Dirac 
point~\cite{Jiang2015:NL,Kong2011:NN,Zhang2011:NC,King2011:PRL,Zhu2011:PRL}.

\begin{figure}[t]
\centering
\includegraphics*[width=8.7cm]{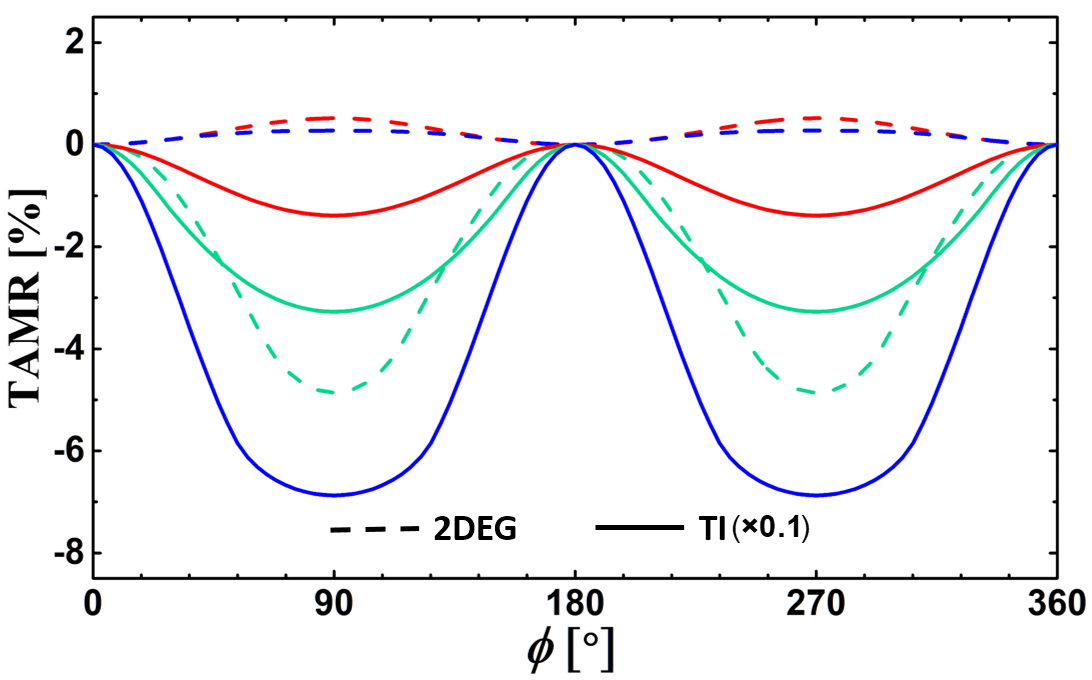}
\caption{Angular dependence of TAMR in 2DEG 
and TI systems with 
$E_F$=10~meV, $V_0$=15~meV, $\Delta=12$~meV (red), 17 meV 
(green), and 26 meV (blue), 
$\alpha=0.093$~eV\AA~(2DEG) and $v_F = 4 \times10^5$  m/s (TI). The  TI values are 10 times larger than labeled on the vertical axis.
}
\label{fig:TAMR_phi}
\end{figure}

TAMR$(\phi)$ of the 2DEG and TI  
for different values of $\Delta$ is represented by dashed and solid lines, respectively, shown in Fig.~\ref{fig:TAMR_phi}. 
For both 2DEG and TI  the maximum TAMR values in Fig.~\ref{fig:TAMR_phi} are quite large compared to  typical values of 
in-plane TAMR $\lesssim 1$ \% in other systems~\cite{Gould2004:PRL,Moser2007:PRL,Park2008:PRL,Uemura2009:APL,Park2008:PRL}. 
This predicted magnitude is particularly striking for a 2DEG with commonly found SOC strength and spin-unpolarized leads.

For $\Delta\ll V_0$, the functional form of the conductance is 
\begin{equation}
G(\phi)\approx A+B\cos^2(\phi),
\label{eq:GAB} 
\end{equation}
where $A$ and $B$ are functions of system parameters other than $\phi$~\cite{Fabian2007:APS}.
It then follows from Eq.~(\ref{eq:tamr-def}) that the angular dependence  
is of the form 
TAMR $\approx B \sin^2(\phi)$, which is precisely the dependence observed in Fig.~\ref{fig:TAMR_phi}. 
Despite the different dispersions of the massive carriers in the 2DEG/F and the massless topological states in the TI/F systems, 
the TAMR in both devices exhibit the same $\sin^2(\phi)$ dependence on $\bm{m}$, preventing their experimental distinction.

\section{VI. Discussion and Conclusions}

While the TAMR$(\phi)$ 
cannot discriminate between the trivial and topological states, we seek
if TAMR can still provide their distinguishing signature. 
Interestingly, from Fig.~\ref{fig:TAMR_phi} we can see that for the 2DEG the sign of $B$ in Eq.~(\ref{eq:GAB}) and even of TAMR can be inverted by changing $\Delta$. 
While TAMR is positive for $\Delta=12$ meV and 26 meV, it becomes negative for $\Delta=17$ meV. 
In contrast, there is no TAMR sign reversal for the TI surface states, shown by solid lines in Fig.~\ref{fig:TAMR_VO_Delta}, which are
computed following the same procedure as in Ref.~\cite{Scharf2016:PRL}.

\begin{figure}[t]
\vspace{-0.3cm}
\centering
\includegraphics*[width=9cm]{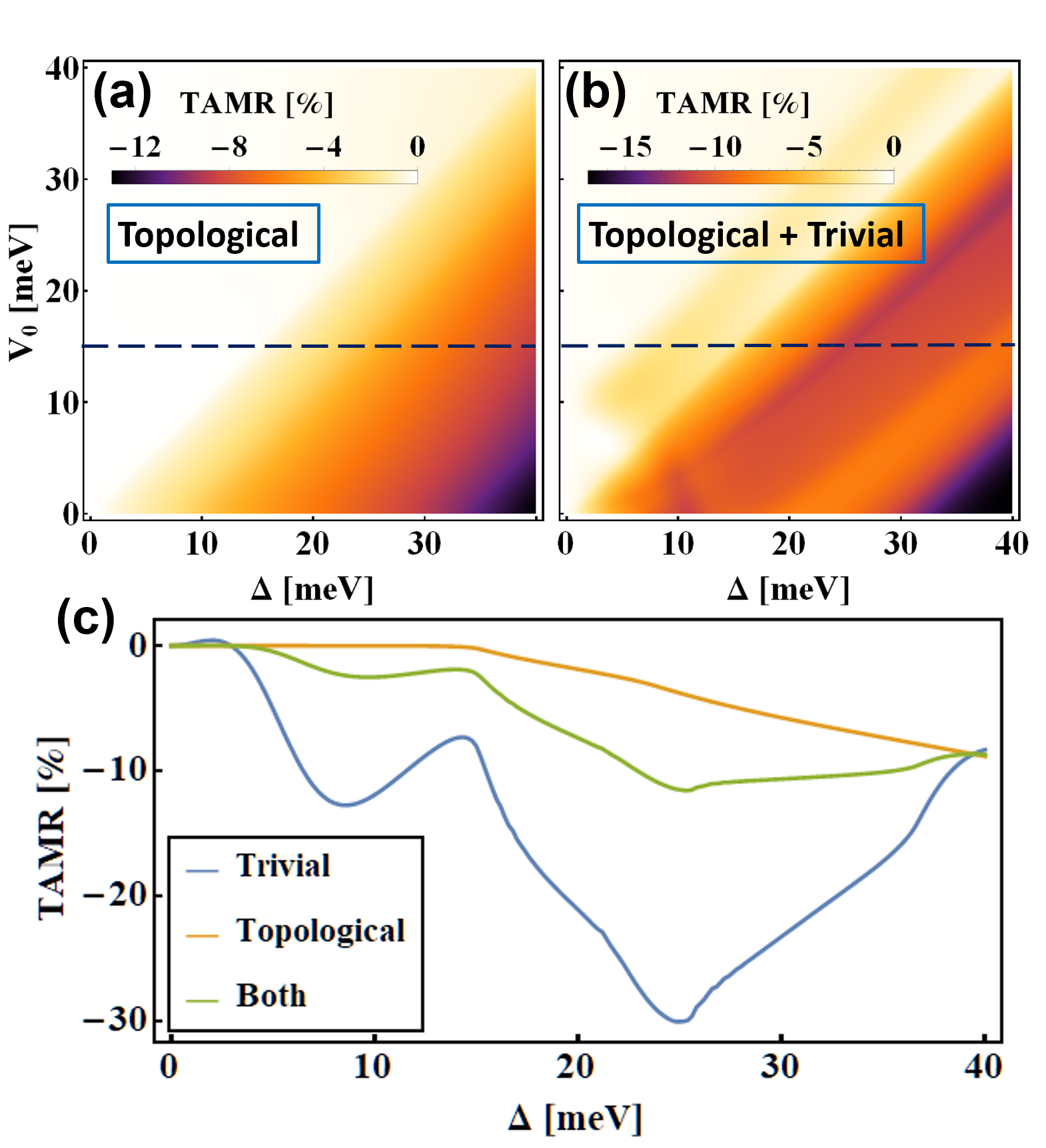}
\caption{ TAMR (a) due to topological surface states in the TI/F system as a function of $V_0$ and $\Delta$ and (b) due to both topological and trivial 
states, where $E_F=10$ meV, $d=13$ nm, $m^\ast=0.27m_0$, and $\alpha=0.36$ eV\AA. (c) Dependence of TAMR amplitude on $\Delta$ with $V_0=15$ meV for different states.}
\label{fig:TAMR_compare}
\end{figure}

Indeed, by comparing TAMR due to (i) only topological states in 
Fig.~\ref{fig:TAMR_compare}(a) and (ii) both topological and trivial states in Fig.~\ref{fig:TAMR_compare}(b),
we can see nonmonotonic TAMR trends in either $\Delta$ or $V_0$ arise only from the  
trivial states. This is better illustrated in Fig.~\ref{fig:TAMR_compare}(c) with TAMR as a function of $\Delta$.  
The main peak of $|$TAMR$|$  at $\Delta=25$ meV originates from the earlier Fermi contour matching argument. 
However, the resonant transmission at $\Delta=9$ meV arises from a different origin of the standing-wave formation in the barrier due to the constructive interference between the two 2DEG/F interfaces (see Appendix~B). 
Such peaks, also resulting from the $\mathcal{P}_{s}\mathcal{T}$ symmetry, have resonant conditions analogous to those
for simple potential barrier systems~\cite{Griffiths:2005}, as analyzed in Appendix~B.

In contrast, with just TI surface states, the Fermi contour shifts the exchange field without changing its diameter due to its linear dispersion relation, and the TAMR lacks such resonance. Therefore, TAMR measurements and their monotonicity in TI/F systems could help to address the 
controversy~\cite{Li2014:NN,Ando2014:NL,Dankert2015:NL,Li2016:NC,Tian2019:NC,Li2019:NC,Li2016:PRB} whether the transport is purely determined by the 
topological states or if there is also a contribution of trivial states.

To realize magnetic proximity effects for the in-plane transport, magnetic insulators are desirable~\cite{Jiang2015:NL,Wei2013:PRL,Lee2013:PRB}. This precludes current flow in the more resistive F region [Fig.~1(a)] and minimizes hybridization with the 2DEG or TI to enable a gate-tunable proximity-induced exchange splitting in their respective states. However, as shown in graphene~\cite{Lazic2016:PRB,Zollner2016:PRB,Xu2018:NC} for tunable magnetic proximity effects one could also employ ferromagnetic metals, separated by an insulating region from the 2DEG or TI. 

While we have focused on a longitudinal transport in a very simple system having no spin-polarized leads, the predicted 
resonant tunneling behavior emerging from a spin-parity-time symmetry of the scattering states is important not just
in explaining a surprisingly large TAMR, 
but also as a sensitive probe  to distinguish between trivial and topological states. The magnitude of TAMR can be further enhanced by 
considering an out-of-plane magnetization, reaching $\sim100\%$ for TIs and expanding possible applications with a single magnetic region ~\cite{Scharf2016:PRL} .  
Our work also 
motivates several direct generalizations which can be explored in a similar geometry, 
including the influence of defects and phonons. 
We expect a rich 
behavior of the transverse response~\cite{Scharf2016:PRL,HuongDang2015:PRB} 
and unexplored resonant Hall effects as well as detecting different states in magnetic topological 
insulators~\cite{Li2015:PRL}. 
The focus on Rashba spin-orbit coupling can be extended in a growing class of van der Waals materials. 
For example, transition metal 
dichalcogenides in addition to their inherent spin-orbit coupling also provide spin-orbit proximity~\cite{Avsar2014:NC,Gmitra2016:PRB,Yan2016:NC,%
Dankert2017:NC,Wang2015:NC,AntonioBenitez2018:APLM,AntonioBenitez2018:NP,AntonioBenitez2020:NM,Douli2020:NL}
and thereby alter spin textures and expected TAMR,
while 2D van der Waals ferromagnets support 
a versatile gate control~\cite{Burch2018:N,Deng2018:N,Xing2017:2DM}. 

\section{Ackowledgments}
This work has been supported by the NSF ECCS-1810266, US ONR N000141712793, and the UB Center for
Computational Research. 

\hspace{0.5cm}
\setcounter{equation}{0}
\renewcommand{\theequation}{A-\arabic{equation}} 
\section*{Appendix A: Analytical solution for \boldmath$\delta$-barrier model}

To investigate tunneling in a two-dimensional electron gas (2DEG) across a single ferromagnetic barrier,
we consider the Hamiltonian of the $\delta$-barrier system is given by
\begin{equation}
\begin{gathered}
H = \frac{p^2}{2{m^\ast}} +\frac{\alpha}{\hbar}\left(\boldsymbol{\sigma}\times\mathbf{p}\right)\cdot\mathbf{\hat{z}} 
+ [V_0 - \Delta (\mathbf{m} \cdot \boldsymbol{\sigma}) ]\, d\, \delta(x), 
\end{gathered}
\label{eq:Hamiltonian_delta}
\end{equation}
where  $m^\ast$ is the effective mass, $\alpha$ is the Rashba spin-orbit coupling (SOC) strength, 
$\mathbf{\hat{z}}$ is the unit vector along the $z$ axis, $\mathbf{p}=(p_x,p_y)$ is the 2D momentum operator, $\boldsymbol{\sigma}$ is the vector of Pauli matrices. $V_0$ describes the potential barrier, and $\Delta$ and $\mathbf{m}$ are the magnitude 
and direction of the proximity-induced ferromagnetic exchange field. The effective barrier width is represented by $d$.

As a linear combination of all possible eigenstates with the same energy and 
the transverse wave vector, $\; k_y$, the scattering states in the left $(l)$ and right $(r)$ 
sides of the barrier can be written as $\psi _\lambda ^{\left( l \right),\left( r \right)}\left( {x,y} \right) =( 1/\sqrt {2S}) {e^{i{k_y}y}}\phi _\lambda ^{\left( l \right),\left( r \right)}\left( x \right)$ with the sample area $S$ and 
\begin{equation}\label{eq:States_left}
\begin{gathered}
\phi _\lambda ^{\left( l \right)}\left( x \right) = \chi _\lambda ^{\left(  +  \right)}{e^{i{k_{x\lambda }}x}} + {r_{\lambda \lambda }}\chi _\lambda ^{\left(  -  \right)}{e^{ - i{k_{x\lambda }}x}} \\ 
+ {r_{\lambda \bar \lambda  }}\chi _{\bar \lambda  }^{\left(  -  \right)}{e^{ - i{k_{x\bar \lambda  }}x}}, \\ 
\end{gathered} 
\end{equation}
\begin{equation}\label{eq:States_right}
\phi _\lambda ^{\left( r \right)}\left( x \right) = {t_{\lambda \lambda }}\chi _\lambda ^{\left(  +  \right)}{e^{i{k_{x\lambda }}x}} + {t_{\lambda \bar \lambda  }}\chi _{\bar \lambda  }^{\left(  +  \right)}{e^{i{k_{x\bar \lambda  }}x}},
\end{equation}
where the helicity $\lambda  =  \pm 1$, $\bar{\lambda}  = -\lambda $, and the spinors are given by 
\begin{equation}
\chi _\lambda ^{\left(  \pm  \right)} = \left( {\begin{array}{*{20}{c}} 1 \\ 
	{ \mp i\lambda {e^{ \pm i{\theta _\lambda }}}} 
	\end{array}} \right), 
\label{eq:spinorA}
\end{equation}	
with ${\theta _\lambda } = \arcsin(k_y/k_\lambda)$. 

With energy and transverse momentum $\hbar k_y$ conservations, we can express the magnitude and $x$ component of the momentum from the two energy bands ($\lambda  =  \pm 1$) as
\begin{equation}
{k_\lambda } =  - \frac{{\lambda \alpha m^\ast}}{{{\hbar ^2}}} + \sqrt {{{\left( {\frac{{\alpha m^\ast}}{{{\hbar ^2}}}} \right)}^2} + \frac{{2m^\ast E}}{{{\hbar ^2}}}}, 
\label{eq:k_E}
\end{equation}
\begin{equation}
{k_{x \lambda}} = \sqrt {{k_\lambda }^2 - k_y^2}.
\label{eq:kx_E_k}
\end{equation}

For a $\delta$ barrier, the boundary conditions are 
\begin{equation}\label{eq:BC_delta}
\begin{gathered}
\phi _\lambda ^{\left( l \right)}\left( {{0^ - }} \right) = \phi _\lambda ^{\left( r \right)}\left( {{0^ + }} \right), \hfill \\
{\left. {d\phi _\lambda ^{\left( r \right)}/dx} \right|_{x = {0^ + }}} - {\left. {d\phi _\lambda ^{\left( l \right)}/dx} \right|_{x = {0^ - }}} \hfill \\
{\text{                     }} = \frac{{2m^\ast d}}{{{\hbar ^2}}}\left[V_0 - \Delta \left(\mathbf{m} \cdot \boldsymbol{\sigma}\right) \right]\phi _\lambda ^{\left( r \right)}\left( {{0^ + }} \right). \hfill \\ 
\end{gathered} 
\end{equation}

By matching the boundary conditions for the scattering states given by Eqs.~\eqref{eq:States_left} and \eqref{eq:States_right}, 
we can obtain the transmission coefficients $t_{\lambda\lambda}$ and $t_{\lambda \bar{\lambda}}$ 
\begin{widetext}
	\begin{equation}\label{eq:Transmission1}
	{t_{\lambda \lambda }} = \frac{{\left( {{B_ + }{C_ - } - {B_ - }{C_ + }} \right)X + \left( {{B_ - }{D_ + } - {B_ + }{D_ - }} \right)Y + \left( {{D_ - }{C_ + } - {D_ + }{C_ - }} \right)Z}}{{\left( {{A_ - }{B_ + } - {A_ + }{B_ - }} \right)X + \left( {{B_ - }{D_ + } - {B_ + }{D_ - }} \right)Y + \left( {{A_ + }{D_ - } - {A_ - }{D_ + }} \right)Z}}, \hfill \\
	\end{equation}
	\begin{equation}\label{eq:Transmission2}
	{t_{\lambda \bar{\lambda}}} = \frac{{\left( {{A_ - }{C_ + } - {A_ + }{C_ - }} \right)X + \left( {{A_ + }{D_ - } - {A_ - }{D_ + }} \right)Y + \left( {{D_ + }{C_ - } - {D_ - }{C_ + }} \right)Y}}{{\left( {{A_ - }{B_ + } - {A_ + }{B_ - }} \right)X + \left( {{B_ - }{D_ + } - {B_ + }{D_ - }} \right)Y + \left( {{A_ + }{D_ - } - {A_ - }{D_ + }} \right)Z}}. \hfill \\
	\end{equation}
\end{widetext}
where 
\begin{equation}\label{eq:A_def}
{A_ \pm } = \left( { - 2{k_{x\lambda }} - i{{\widetilde V}_0}} \right)f_{ \pm \lambda ,\lambda }^{\left(  +  \right)} + i\left( {\widetilde {\bm{\Delta}}  \cdot {\bm{s}_{ \pm \lambda ,\lambda }}} \right),
\end{equation}
\begin{equation}\label{eq:B_def}
{B_ \pm } = \left( { - {{k_{x\lambda }} - {k_{x\bar \lambda  }}} - i{{\widetilde V}_0}} \right)f_{ \pm \lambda ,\bar \lambda  }^{\left(  +  \right)} + i\left( {\widetilde {\bm{\Delta}}  \cdot {\bm{s}_{ \pm \lambda ,\bar \lambda  }}} \right),
\end{equation}
\begin{equation}\label{eq:C_def}
{C_ \pm } =  - 2{k_{x\lambda }}f_{ \pm \lambda ,\lambda }^{\left(  +  \right)},
\end{equation}
\begin{equation}\label{eq:D_def}
{D_ \pm } = \left( {{k_{x\lambda }} - {k_{x\bar \lambda  }}} \right)f_{ \pm \lambda ,\bar \lambda  }^{\left(  -  \right)},
\end{equation}
\begin{equation}\label{eq:X_def}
X = f_{\lambda \bar \lambda  }^{\left(  -  \right)}sf_{\bar \lambda  \lambda }^{\left(  -  \right)} - f_{\bar \lambda  \bar \lambda  }^{\left(  -  \right)}f_{\lambda \lambda }^{\left(  -  \right)},
\end{equation}
\begin{equation}\label{eq:Y_def}
Y =  - 2f_{\bar \lambda  \lambda }^{\left(  -  \right)} + f_{\bar \lambda  \lambda }^{\left(  +  \right)}f_{\lambda \lambda }^{\left(  -  \right)},
\end{equation}
\begin{equation}\label{eq:Z_def}
Z =  - f_{\lambda \bar \lambda  }^{\left(  +  \right)}f_{\bar \lambda  \lambda }^{\left(  -  \right)} + 2f_{\lambda \lambda }^{\left(  -  \right)},
\end{equation}
with
\begin{equation}\label{eq:f_def}
f_{\lambda ,\lambda '}^{\left(  \pm  \right)} = \chi {_\lambda ^{\left(  +  \right)\dag} }\chi _{\lambda '}^{\left(  \pm  \right)},
\end{equation}
\begin{equation}\label{eq:s_def}
{\bm{s}_{\lambda ,\lambda '}} = \chi {_\lambda ^{\left(  +  \right)\dag} }\bm{\sigma} \chi _{\lambda '}^{\left(  +  \right)},  
\end{equation}
and $\widetilde{\bm{\Delta}}  = (2m^\ast d /\hbar ^2)\Delta\mathbf{m}$, $\widetilde{V_0} = (2m^\ast d /\hbar ^2) V_0$.
In particular, for normal incidence the transmission coefficients reduce to 
\begin{equation}\label{eq:Normal_T1}
{t_{\lambda \lambda }} = \frac{{\left( {{k_\lambda } + {k_{ \bar{\lambda} }}} \right)\left[ {\left( {{k_\lambda } + {k_{ \bar{\lambda} }}} \right) + i{\widetilde{V_0}} - i\widetilde{\Delta} \lambda \sin \phi } \right]}}{{{{\left[ {\left( {{k_\lambda } + {k_{ \bar{\lambda} }}} \right) + i{\widetilde{V_0}}} \right]}^2} + {\widetilde{\Delta} ^2}}},
\end{equation}
\begin{equation}\label{eq:Normal_T2}
{t_{\lambda \bar{\lambda} }} = \frac{{\lambda \left( {{k_\lambda } + {k_{ \bar{\lambda} }}} \right)\widetilde{\Delta} \cos \phi }}{{{{\left[ {\left( {{k_\lambda } + {k_{\bar{\lambda}}}} \right) + i{\widetilde{V_0}}} \right]}^2} + {\widetilde{\Delta} ^2}}},
\end{equation}
where  $\widetilde{\Delta} = | \widetilde{\bm{\Delta}} | $ and $\phi$ is the angle between magnetization 
${\mathbf M}$ (or, equivalently, ${\mathbf m}$) 
and the $+x$ axis.

The $\delta$-barrier model agrees with the square-barrier model in the angular dependence and the resonances around $\Delta=V_0$, 
but it fails to show the resonance  originating from the constructive interference, recall Fig.~\ref{fig:TAMR_compare}. 
Therefore, we will only use this model to analyze the angular dependence.

\setcounter{equation}{0}
\renewcommand{\theequation}{B-\arabic{equation}} 
\section*{Appendix B: Transmission Resonances}
\subsection{Symmetry analysis}

\begin{figure}[h]
\centering
\includegraphics*[width=9cm]{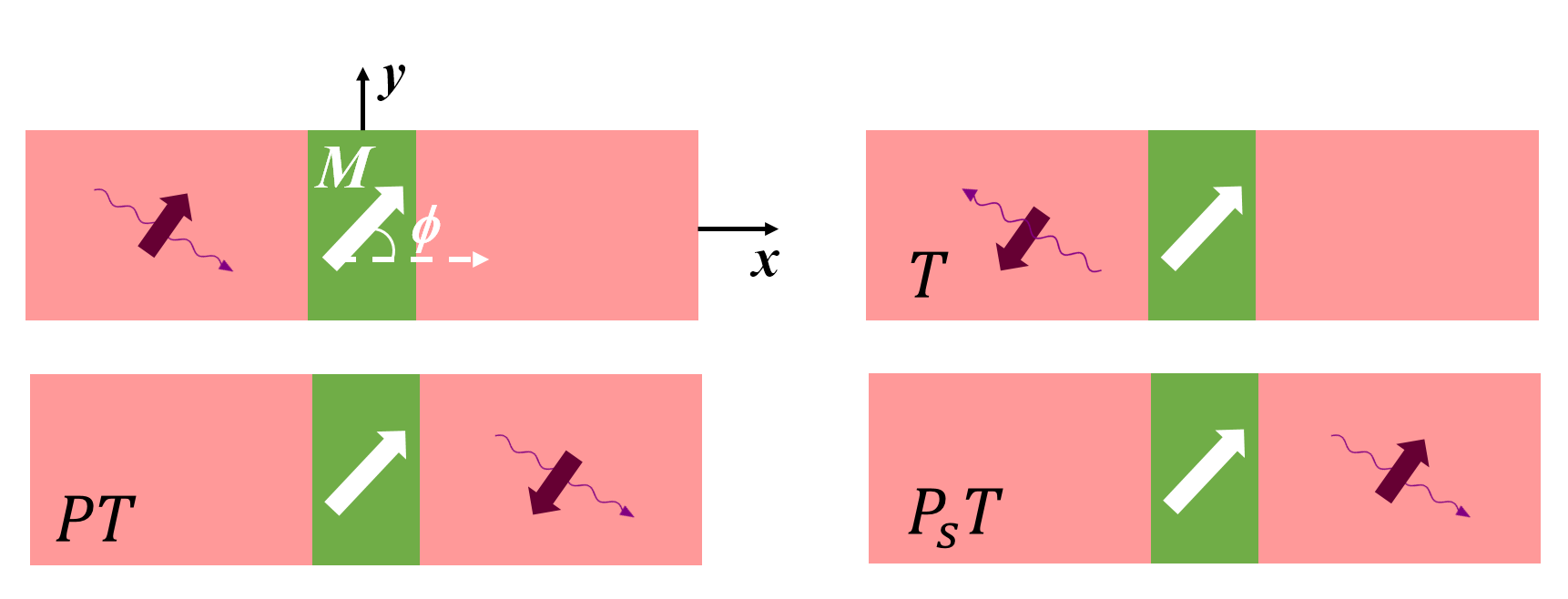}
\caption{Spin-parity-time symmetry of the scattering states. Successive action of several discrete symmetry
operations on an incident wave with an in-plane spin (black arrow) on the left side of the magnetic barrier. $\mathcal{T}$ is the time reversal, 
$\mathcal{P}$ the space inversion, and $\mathcal{P}_s \equiv \mathcal{P}\sigma_z$, where $\sigma_z$ is the Pauli matrix,
inverts both the spin and the position. By applying the $\mathcal{P}_s\mathcal{T}$ operator the incident wave on the left is transformed to itself, but as a transmitted wave on the right. }
\label{fig:PST}
\end{figure}

In simple spinless barrier systems the resonant conditions for a perfect transparency with transmission $T=1$ are well understood~\cite{Griffiths:2005}.
However, much less is known how to generalize those conditions for spinful systems with magnetic barriers. Following the arguments to generalize
such conditions of resonant transmission for scattering states outlined in the Introduction we can complement them with the illustration in Fig.~\ref{fig:PST}.

We can then intuitively understand that the resonant transmission occurs when the scattering states are invariant (up to a phase difference) 
under the operation 
\begin{equation}
\mathcal{P}_s\mathcal{T} \equiv \mathcal{P}{\sigma _z}\mathcal{T},
\label{eq:PST}
\end{equation}
where
$\mathcal{P}$ is the parity operator in the 
$x$ direction, $\mathcal{T} =  - i{\sigma _y}\mathcal{K}$ is the time-reversal operator, and $\mathcal{K}$ is the complex 
conjugation operator. While $\mathcal{P}{\sigma _z}\mathcal{T}$ does not commute with $H$  
in Eqs.~(\ref{eq:Hamiltonian}) or (\ref{eq:Hamiltonian_delta}), such symmetry is satisfied at resonances. 
As shown in Fig.~\ref{fig:PST} by applying the $\mathcal{P}_s\mathcal{T}$ operator the incident wave on the left is transformed to itself but as a transmitted wave on the right. Therefore, scattering states which are eigenfunctions of $\mathcal{P}_s\mathcal{T}$ experience perfect transmission. 
At the end of the appendix we show how for 
strong magnetic proximity this case of a generalized transmission
resonance is reduced to the well-known condition of spinless systems.

According to the wave functions 
of the scattering states in 
Eq.~(\ref{eq:States_finite_barrier}), there are four eigenstates in the lead with $x$-component wave vectors 
$\pm k_{x\lambda},\pm k_{x \bar\lambda}$. The $x$-component wave vectors for the lead states from the inner Fermi contour $\pm k_{x\bar\lambda}$ can be purely imaginary when the incident particles are from the outer Fermi contour ($\lambda = -1$) with incident angles greater than the critical angle $\theta_0$. Therefore, we will discuss the resonance conditions for these two situations ($\pm k_{x\bar\lambda}$ real or purely imaginary) separately.

\subsubsection{Case 1: 
Four Propagating Lead Eigenstates}
\label{SubSec:4_lead_states}
From  the particle current conservation, a 
perfect transmission requires no reflected current. If all the eigenstates 
in the leads are propagating, a 
perfect transmission means that no reflected waves should exist.
In this situation, the scattering states can be written as
\begin{equation}\label{eq:4_lead_states}
{\phi _\lambda }\left( x \right) = \left\{ \begin{array}{ll}
{e^{i{k_{x\lambda }}x}}\chi _\lambda ^{\left(  +  \right)}, & x < -d/2, \hfill \\
\sum\limits_{\lambda ' =  \pm 1} {\sum\limits_n {A_{\lambda \lambda '}^{\left( n \right)}{e^{i\widetilde k_{x\lambda '}^{(n)}x}}\tilde \chi _{\lambda '}^{\left( n \right)}} }, & -d/2 < x < d/2, \hfill \\
{e^{i\eta }}{e^{i{k_{x\lambda }}x}}\chi _\lambda ^{\left(  +  \right)}, & x > d/2, \hfill \\ 
\end{array}\right.
\end{equation}
where $\eta$ is an arbitrary phase. 
The continuity equations of the wave function and its derivative at $x = -d/2$ 
lead to the following system of four linear equations,
\begin{equation}
\sum\limits_{j = 1}^4 {{M_{ij}}{A_j}}  = {b_i},\, i=1,2,3,4,
\label{eq:linear_eqs_left}
\end{equation}
where $M_{ij}$ are elements of the matrix of the system of linear equations, and the coefficients of the barrier states $A_{\lambda \lambda '}^{\left( n \right)}$ are written in the form of $A_j$. Except for normal incidence when $\mathbf{m}\parallel\mathbf{y}$, all equations
are linearly independent, so each $A_j$ is uniquely determined by these linear equations. On the other hand, the boundary conditions at $x = d/2$ can also be written as 
\begin{equation}\label{eq:linear_eqs_right}
\sum\limits_{j = 1}^4 {{M_{ij}}{A_j}{e^{i{{\tilde k}_j}d}}}  = {b_i}{e^{i\left( {\eta  + {k_{x\lambda }}d} \right)}},\, i=1,2,3,4,
\end{equation}
where the $x$ components of the barrier wave vectors are written in the form of $\widetilde k_j$ since there is no need to distinguish which energy band they are from. 
These boundary conditions are satisfied when and only when
\begin{equation}
{e^{i{{\widetilde k}_j}d}} = {e^{i\left( {\eta  + {k_{x\lambda }}d} \right)}},\, j=1,2,3,4,
\label{eq:4_condition}
\end{equation}
is valid for all the barrier states. If there exist decaying barrier states, Eq.~\eqref{eq:4_condition} cannot be fulfilled. If all the barrier states are propagating, this condition can be satisfied when
\begin{equation}\label{eq:4_condition1}
\left( {{{\widetilde k}_j} - {{\widetilde k}_1}} \right)d = 2\pi{n_j},\, j=2,3,4,
\end{equation}
where $n_j$ are any integers. Since all the propagating eigenstates are invariant (up to a phase difference) under $\mathcal{P}{\sigma _z}\mathcal{T}$ (see next subsection), the $\mathcal{P}{\sigma _z}\mathcal{T}$ symmetry is always fulfilled for the resonant states in this case.
However, for the system here considered, ${{\widetilde k}_j}d \sim 10$ and
Eq.~\eqref{eq:4_condition1} can hardly be satisfied. 

The normal incidence when $\mathbf{m}\parallel\mathbf{y}$ is a special case, where the spins of all scattering states are parallel to each other, which makes the system ``spinless" and perfect transmission becomes possible. In summary, except for a few special cases, the resonance will not happen when all the lead states are propagating.

\subsubsection{Case 2: Two Propagating and Two Decaying Lead Eigenstates}\label{SubSec:2and2_lead states}
If there are decaying lead eigenstates, they can exist in the reflected waves at perfect transmission since they do not carry any current. This will happen when the incidence is from the $\lambda=-1$ band and the incident angle is greater than the critical angle $\theta_0$. In this situation, the scattering states are given by
\begin{equation}\label{eq:2_lead_states}
{\phi _\lambda }\left( x \right) = \left\{ {\begin{array}{*{20}{l}}
	{{e^{i{k_{x\lambda }}x}}\chi _\lambda ^{\left(  +  \right)} + r{e^{\left| {{k_{x\bar \lambda }}} \right|x}}\chi _{\bar \lambda }^{\left(  -  \right)},}&{x <  - d/2,} \\ 
	{\sum\limits_{\lambda ' =  \pm 1} {\sum\limits_n {A_{\lambda \lambda '}^{\left( n \right)}{e^{i\widetilde k_{x\lambda '}^{(n)}x}}\tilde \chi _{\lambda '}^{\left( n \right)}} } ,}&{ - d/2 < x < d/2,} \\ 
	{{e^{i{\eta _0}}}{e^{i{k_{x\lambda }}x}}\chi _\lambda ^{\left(  +  \right)} + t{e^{ - \left| {{k_{x\bar \lambda }}} \right|x}}\chi _{\bar \lambda }^{\left(  +  \right)},}&{x > d/2.} 
	\end{array}} \right.
\end{equation}

\begin{figure}[b]
\vspace{-0.4cm}
\centering
\includegraphics*[width=8cm]{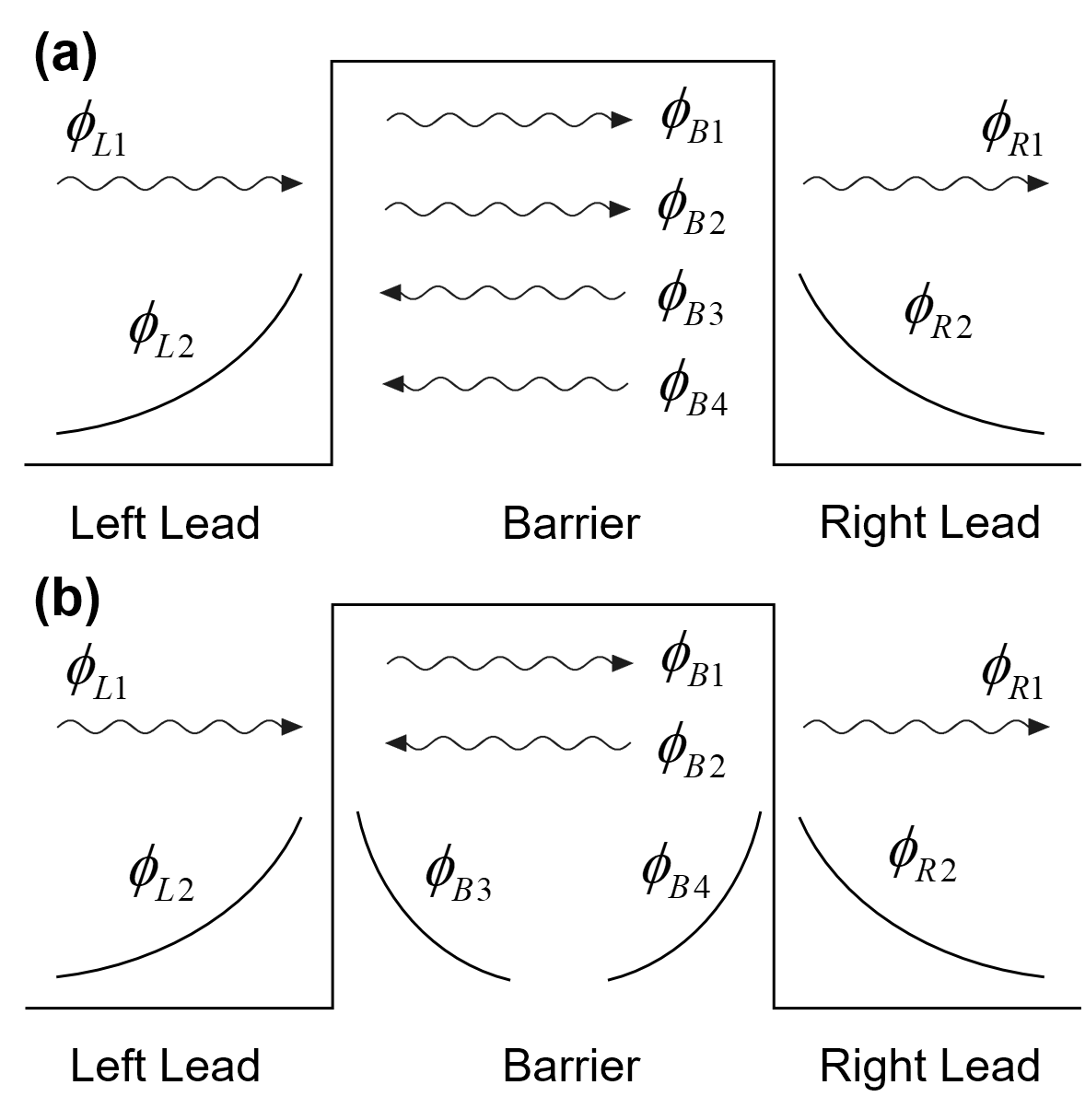}
\caption{Schematic of the 
two propagating and two decaying lead eigenstates for (a) four and (b) two propagating states in the barrier.
Under the  symmetry $\mathcal{P}{\sigma _z}\mathcal{T}$ operation, up to a phase difference, the propagating states 
inside the barrier remain the same, while $\phi_{L1}$ becomes $\phi_{R1}$ and the decaying states become their partners ($\phi_{L2}$ to $\phi_{R2}$ and $\phi_{B3}$ to $\phi_{B4}$).} 
\label{fig:2&4_barrier_states}
\end{figure}

Similar as illustrated in Fig.~\ref{fig:2&4_barrier_states}, the spatial 
parts of the wave functions of the propagating states ( $\sim {e^{ikx}}$) remain the same (up to a phase difference) under the $\mathcal{P}{\sigma _z}\mathcal{T}$ operation, while those of the decaying states ($ \sim {e^{ \pm \left| k \right|x}}$) change to $ \sim {e^{ \mp \left| k \right|x}}$.
For the spinors of the propagating lead eigenstates, they can be written as ${\chi ^{\left(  \pm  \right)}} = \left( {\begin{array}{*{20}{c}}
	1 \\ 
	{{e^{ \pm i\beta }}} 
	\end{array}} \right)$, with $\beta  = {\theta _{\lambda  =  - 1}} + \pi /2$. The spinors  
	 are in the $xy$ plane and satisfy the following relation under the operation $\mathcal{P}{\sigma _z}\mathcal{T}$ ($\mathcal{P}$ has no impact on the spinor),
\begin{equation}\label{eq:propagating_symmetry}
{\sigma _z}\mathcal{T}{\chi ^{\left(  \pm  \right)}} = - {e^{ \mp i\beta }}{\chi ^{\left(  \pm  \right)}}.
\end{equation}
For the decaying states, whose spinors can be expressed by $\chi {'^{\left(  -  \right)}} = \left( {\begin{array}{*{20}{c}}
	{{u^*}} \\ 
	1 
	\end{array}} \right),\chi {'^{\left(  +  \right)}} = \left( {\begin{array}{*{20}{c}}
	1 \\ 
	u 
	\end{array}} \right)$ with $u = {u^\ast} = (\left| {k_{x,\lambda  = 1}} \right| + k_y)/ {\sqrt {{k_y^2 - \left| k_{x,\lambda  = 1} \right|}^2}}$, their spins point out of the $xy$ plane, 
	and the relation becomes
\begin{equation}
{\sigma _z}\mathcal{T}\chi {'^{\left(  \pm  \right)}} = - \chi {'^{\left(  \mp  \right)}}.
\label{eq:decaying_symmetry}
\end{equation}
Therefore, if $t$ from Eq.~(\ref{eq:2_lead_states}) 
is set to be 
\begin{equation}
t = r{e^{i\left( {\beta  + {\eta _0} - 2{\text{Arg}}\left( r \right)} \right)}},
\label{eq:t}
\end{equation}
the scattering states in the lead have the following symmetry 
\begin{equation}
{\phi _R}\left( x \right) = {e^{i\eta_L }}\mathcal{P}{\sigma _z}\mathcal{T}{\phi _L}\left( x \right),
\label{eq:lead_symmetry}
\end{equation}
where $\eta_L=\eta_0 + \beta - \pi$.

We will show that the barrier eigenstates can also satisfy this symmetry so that the boundary conditions can be fulfilled at both $x = -d/2$ and $x = d/2$.

If all the barrier eigenstates are propagating [Fig.~\ref{fig:2&4_barrier_states}(a)], their spinors can always be written as $\widetilde\chi_i = \left( {\begin{array}{*{20}{c}}
	1 \\ 
	{{e^{i\gamma_i }}} 
	\end{array}} \right)$. 
Since no matter which energy band the states belong to, they share the same properties in spinors and wave vectors, we do not need to distinguish $\lambda ' =  \pm 1$. The scattering states in the barrier can then be written as
\begin{equation}\label{4 propagating barrier states}
{\phi _B}\left( x \right) = \sum\limits_{i = 1}^4 {{A_i}{e^{i{{\widetilde k}_i}x}}{{\tilde \chi }_i}}.
\end{equation}
We notice that for the spinor of a propagating barrier eigenstate,  ${\sigma _z}\mathcal{T}{{{\tilde \chi }_i}}  = - {e^{-i\gamma_i }}{{{\tilde \chi }_i}}.$ 
Applying the operation $\mathcal{P}{\sigma _z}\mathcal{T}$ on the barrier states, we have 
\begin{equation}
\mathcal{P}{\sigma _z}\mathcal{T}{\phi _B}\left( x \right) = \sum\limits_{i = 1}^4 {{A_i}{e^{ - i\left( {2{\text{Arg}}\left( {{A_i}} \right) + {\gamma _i} - \pi} \right)}}{e^{i{{\widetilde k}_i}x}}{{\tilde \chi }_i}}.
\label{eq:4_states_transformed}
\end{equation}
The phase difference arising from the transformation for each term can be written as
\begin{equation}
{\eta _i} = 2{\text{Arg}}\left( {{A_i}} \right) + {\gamma _i} -\pi.
\label{eq:transform_phase}
\end{equation}
The resonance will occur when and only when all the phase differences are equal or have difference of $2n\pi$, i.e.,
\begin{equation}\label{eq:4_barrier_resonance_condition}
{\eta _1}+ 2n_1 \pi={\eta _2}+ 2n_2 \pi={\eta _3}+ 2n_3 \pi={\eta _4},
\end{equation}
where $n_i$ are integers.

If there are two decaying states and two propagating states in the barrier  [Fig.~\ref{fig:2&4_barrier_states}(b)], the scattering states in the barrier can be written as
\begin{equation}
{\phi _B}\left( x \right) = \sum\limits_{i = 1}^2 {{A_i}{e^{i{{\widetilde k}_i}x}}{{\tilde \chi }_i}}  + \sum\limits_{i = 1}^2 {{B_i}{e^{i\widetilde k{'_i}x}}\tilde \chi {'_i}},
\label{eq:2_barrier_states}
\end{equation}
where the spinors of the first two propagating states are given by  ${{\tilde \chi }_i} = \left( {\begin{array}{*{20}{c}}
	1 \\ 
	{{e^{i{\gamma _i}}}} 
	\end{array}} \right)$, with ${\gamma _i} = \tilde \theta _{\lambda ' =  - 1}^{\left( i \right)} + \pi /2$. 
	We recall that $\tilde \theta _{\lambda ' =  - 1}^{\left( i \right)}$ is defined in Sec.~II from Eq.~(\ref{eq:barrier_spinor}) and the spinors of the last two decaying ones are given by ${{\tilde \chi }_1}' = \left( {\begin{array}{*{20}{c}}
	1 \\ 
	w 
	\end{array}} \right),{{\tilde \chi }_2}' = \left( {\begin{array}{*{20}{c}}
	{{w^*}} \\ 
	1 
	\end{array}} \right)$, 
	with $w = [ - i ( {\tilde k{'_1} + {\Delta _y}/\alpha } ) + ( k_y - \Delta _x/\alpha)]/{\sqrt {{ ( {\tilde k{'_1}} + \Delta _y/\alpha)^2} + ( k_y - \Delta _x/\alpha)^2}}$.
For the decaying states their $x$ component of the wave vectors 
are complex conjugates to each other. 
Thus, assuming ${\widetilde k_{\operatorname{Im} }}'>0$, 
we can write them in the following form 
\begin{equation}
\widetilde k{'_{1,2}} = {\widetilde k_{\operatorname{Re} }}' \mp i{\widetilde k_{\operatorname{Im} }}'.
\label{eq:decaying_k} 
\end{equation}
Similar to the decaying lead states, the spinors of the decaying barrier states $\widetilde \chi {'_i}$ also have the following relation under the operation $\mathcal{P}{\sigma _z}\mathcal{T}$, 
\begin{equation}\label{eq:decaying_symmetry}
{\sigma _z}\mathcal{T}\widetilde \chi {'_{1,2}} = - \widetilde \chi {'_{2,1}}.
\end{equation}
Applying the operation $\mathcal{P}{\sigma _z}\mathcal{T}$ on the barrier states, we obtain  
\begin{equation}\label{eq:2_barrier_states_transformed}
\begin{gathered}
\mathcal{P}{\sigma _z}\mathcal{T}{\phi _B}\left( x \right) = \sum\limits_{i = 1}^2 {{A_i}{e^{ - i\left( {2{\text{Arg}}\left( {{A_i}} \right) + {\gamma _i} -\pi} \right)}}{e^{i{{\widetilde k}_i}x}}{{\tilde \chi }_i}}  \hfill \\
+ {e^{ - i\left[ {{\text{Arg}}\left( {{B_1}} \right) + {\text{Arg}}\left( {{B_2}} \right)} - \pi \right]}}{e^i}^{\widetilde k{'_{\operatorname{Re} }}x} \hfill \\
\times \left( {\left| {\frac{{{B_2}}}{{{B_1}}}} \right|{B_1}{e^{\widetilde k{'_{\operatorname{Im} }}x}}\tilde \chi {'_1} + \left| {\frac{{{B_1}}}{{{B_2}}}} \right|{B_2}{e^{ - \widetilde k{'_{\operatorname{Im} }}x}}\tilde \chi {'_2}} \right). 
 \hfill \\ 
\end{gathered} 
\end{equation}
The resonance will occur when and only when
\begin{equation}\label{eq:2_barrier_resonance_condition}
\left\{ \begin{gathered}
2{\text{Arg}}\left( {{A_1}} \right) + {\gamma _1} = 2{\text{Arg}}\left( {{A_2}} \right) + {\gamma _2} + 2n\pi,  \hfill \\
2{\text{Arg}}\left( {{A_1}} \right) + {\gamma _1} = {\text{Arg}}\left( {{B_1}} \right) + {\text{Arg}}\left( {{B_2}} \right) + 2m\pi,  \hfill \\
\left| {{B_1}} \right| = \left| {{B_2}} \right|, \hfill \\ 
\end{gathered}  \right.
\end{equation}
where $m,n$ are integers. These resonance conditions show that the two propagating and two decaying states satisfy the symmetry independently and finally their phase changes are required to be the same. As a result, the scattering states in the barrier satisfies the following symmetry 
\begin{equation}\label{eq:barrier_symmetry}
{\phi _B}\left( x \right) = {e^{i\eta_B }}\mathcal{P}{\sigma _z}\mathcal{T}{\phi _B}\left( x \right),
\end{equation}
where $\eta_B  = {\text{Arg}}\left( {{B_1}} \right) + {\text{Arg}}\left( {{B_2}} \right) -\pi + 2n\pi$ with a certain integer $n$ that makes $\eta_B \in \left[ {0,2\pi } \right]$. Under the operation $\mathcal{P}{\sigma _z}\mathcal{T}$, the phase shift of barrier and lead states are required to be the same, i.e., ${\eta _L} = {\eta _B} \equiv \eta $, which determines ${\eta _0} = {\text{Arg}}\left( {{B_1}} \right) + {\text{Arg}}\left( {{B_2}} \right) + 2n\pi - \beta $.

By combining  Eqs.~\eqref{eq:lead_symmetry} and
~\eqref{eq:barrier_symmetry}, 
we have the symmetry for the whole scattering states 
\begin{equation}\label{eq:whole_symmetry}
\phi \left( x \right) = {e^{i\eta }}\mathcal{P}{\sigma _z}\mathcal{T}\phi \left( x \right) = {e^{i\eta }}{\sigma _z}\mathcal{T}\phi \left( { - x} \right),
\end{equation}
while for its first derivative it follows 
\begin{equation}\label{eq:derivative_symmetry}
\phi '\left( x \right) =  - {e^{i\eta }}\mathcal{P}{\sigma _z}\mathcal{T}\phi '\left( x \right) =  - {e^{i\eta }}{\sigma _z}\mathcal{T}\phi '\left( { - x} \right).
\end{equation}
These symmetry requirements will add three more constraints to the system of three real equations: 
Eq.~\eqref{eq:4_barrier_resonance_condition} for four propagating barrier eigenstates or Eq.~\eqref{eq:2_barrier_resonance_condition} 
for two propagating and two decaying barrier eigenstates. 
Together with the boundary conditions at $x = -d/2$, which are four complex equations or equivalently eight real equations, we are able to solve for the  total of five complex variables (reflection $r$ and four coefficients for the barrier states) and one real system parameter for resonances. 
For example, we may find out the magnitude of the proximity-induced exchange field $\Delta$ for the 
resonance with fixed Fermi energy $E_F$, barrier potential $V_0$, and barrier width $d$. At $x = d/2$, since the barrier and lead states both satisfy the symmetry, the boundary conditions will be fulfilled automatically. Now we can conclude  
that the system at higher-order resonances does obey the symmetry $\mathcal{P}{\sigma _z}\mathcal{T}$ and the conditions given by Eq.~\eqref{eq:4_barrier_resonance_condition} or Eq.~\eqref{eq:2_barrier_resonance_condition} are the condition for these resonances.

\subsection{Resonances due to interference}
\label{SubSec:Resonance due to interference}
\begin{figure}[b]
	\centering
	\includegraphics*[trim=0.1cm 0.2cm 0.2cm 0.1cm,clip,width=8.5cm]{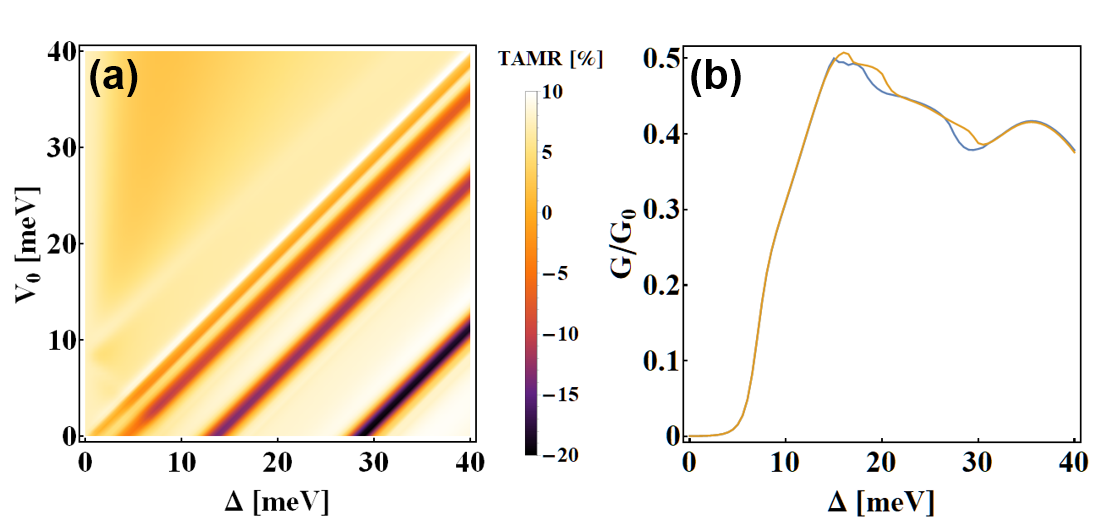}
	\caption{(a) Dependence of the 
	TAMR amplitude on $V_0$ and $\Delta$ for 2DEG system with a $d=50$~nm thick barrier, $E_F=10$~meV, and $\alpha=0.093$~eV\AA. (b) The corresponding conductances with fixed $V_0=15$~meV, where blue and yellow curves denote conductance for $\mathbf{m} \parallel \mathbf{x}$ and $\mathbf{m} \parallel \mathbf{y}$, respectively.}
\label{fig:other_resonances}
\end{figure}
In the square barrier model, when the barrier width is increased to 50 nm, we can see
in Fig.~\ref{fig:other_resonances} 
more than one resonance for the transmission. The one near $\Delta = V_0$ is due to the barrier contour matching. The others arise from the formation of standing de Broglie waves in the barrier due to constructive interference. The resonance condition is $\left({\tilde k}_2-{\tilde k}_1\right)d = 2n\pi-\delta$, where $n=1,2,3\ldots$, ${\tilde k}_{1,2}$ are the $x$ component of the propagating wave vectors in the barrier (assuming ${\tilde k}_2>{\tilde k}_1$), $d$ is the width of the barrier, $\delta$ is a correction proportional to $1/\sqrt \Delta$ (see next subsection). Since the higher-order resonances occur at lager 
$\Delta$, the magnitude of the propagating barrier wave vector is much larger than $k_y$. Therefore, the $x$ component of the propagating barrier wave vector is almost the same for all the incoming states with different incident angles, which means the transmission resonances occur for all the incoming states simultaneously when the resonance condition is satisfied and thus the maximum conductance is reached.

In the limit of $\alpha \to 0$, we have the maximum condition 
\begin{equation}
\Delta  = {V_0} - {E_F} + \frac{\pi ^2 \hbar ^2 n^2}{2m^\ast d ^2},
\label{eq:Delta_max}
\end{equation}
 for both $G(0)$ and $G(\pi/2)$. However, when $\alpha \ne0$, ${{\tilde k}_2 - {\tilde k}_1}$ varies as   
${\mathbf M}$ is rotated, 
which means the maxima for $G(0)$ and $G(\pi/2)$ will be achieved at different $\Delta$. Therefore, similar to the situation near $V_0=\Delta$, a resonant TAMR will arise from the small difference in the peak conditions. With fixed $E_F$ and $V_0$, up to the lowest order correction of $\alpha$, we can derive the difference in $\Delta$ at the same order maxima of $G(0)$ and $G(\pi/2)$
\begin{equation}\label{higher order resonance peak shift}
\Delta_\text{diff}  = \frac{{ {\pi ^2}{\hbar ^2}{n^2}/({2m^\ast d ^2}) - {V_0} + {E_F}}}{{ {\pi ^2}{\hbar ^2}{n^2}/({2m^\ast d ^2}) + {V_0} - {E_F}}} \times \frac{{{\alpha ^2}m^\ast}}{{2{\hbar ^2}}}.
\end{equation}

\subsection{Resonances in the strong magnetic proximity limit}
\label{SubSec:Resonances in the strong magnetization limit}

The higher-order resonances with large 
$\Delta$ in Fig.~\ref{fig:other_resonances} are in the regime of Case 2, with two propagating and two decaying barrier states. The barrier wave function is given by 
Eq.~\eqref{eq:2_barrier_states}.
The wave functions at resonance satisfy the symmetry $\mathcal{P}{\sigma _z}\mathcal{T}$ and the resonance condition Eq.~\eqref{eq:2_barrier_resonance_condition}.
On the other hand, such resonances arising from the formation of standing waves require the phase of the propagating waves to remain
 unchanged after one cycle of travel in the barrier. The total phase change during one cycle comes from the propagation of the waves and the phase changes due to the reflections at the interfaces. The phase change coming from the propagation is ${\varphi _1} = {{\tilde k}_2}d - {{\tilde k}_1}d$. The phase change at the interfaces ${\varphi _2}$ can be estimated by considering 
the scattering at one interface while ignoring the other. By treating the left-moving (right-moving) barrier propagating state as the incident state at the left (right) interface, we can write the scattering states of such single-interface models as
\begin{equation}\label{eq:interface_wave_function1}
{\phi _{s1}}\left( x \right) = \left\{ {\begin{array}{*{20}{c}}
	{{t_{s1}}{e^{ - i{k_1}x}}{\chi ^{\left(  -  \right)}} + {t_{s1}}'{e^{\left| {{k_2}} \right|x}}\chi {'^{\left(  -  \right)}},}&{x < 0,} \\ 
	{{e^{i{{\widetilde k}_1}x}}{{\tilde \chi }_1} + {r_{s1}}{e^{i{{\widetilde k}_2}x}}{{\tilde \chi }_2} + {r_{s1}}'{e^{i\widetilde k{'_2}x}}\tilde \chi {'_2},}&{x > 0} 
	\end{array}} \right.
\end{equation}
for the left interface and
\begin{equation}\label{eq:interface_wave_function2}
{\phi _{s2}}\left( x \right) = \left\{ {\begin{array}{*{20}{c}}
	{{e^{i{{\widetilde k}_2}x}}{{\tilde \chi }_2} + {r_{s2}}{e^{i{{\widetilde k}_1}x}}{{\tilde \chi }_1} + {r_{s2}}'{e^{i\widetilde k{'_1}x}}\tilde \chi {'_1},}&{x < 0,} \\ 
	{{t_{s2}}{e^{i{k_1}x}}{\chi ^{\left(  +  \right)}} + {t_{s2}}'{e^{ - \left| {{k_2}} \right|x}}\chi {'^{\left(  +  \right)}},}&{x > 0} 
	\end{array}} \right.
\end{equation}
for the right interface, where ${k_1} \equiv {k_{x,\lambda  =  - 1}}$ (${k_2} \equiv {k_{x,\lambda  = 1}}$) is the 
$x$-component wave vector of the propagating (decaying) lead states (defined in Appendix A), 
${{\tilde k}_{1,2}}$ ($\tilde k{'_{1,2}}$) are the $x$-component wave vectors of the propagating (decaying) barrier states [assuming ${{\tilde k}_1} < {{\tilde k}_2}$ and $\operatorname{Im} ( {{{\tilde k}_1}'}) < \operatorname{Im} ( {{{\tilde k}_2}'})$]. The spinors of the lead states are defined above Eqs.~\eqref{eq:propagating_symmetry} and ~\eqref{eq:decaying_symmetry}, while the barrier spinors are the same as those defined in Eq.~\eqref{eq:2_barrier_states}.
The total phase change due to the reflections at the two interfaces is given by
\begin{equation}\label{eq:total_phase_change at the interfaces}
{\varphi _2} = {\text{Arg}}\left( {{r_{s1}}} \right) + {\text{Arg}}\left( {{r_{s2}}} \right).
\end{equation}
For clarity, we consider the large $\Delta$ limit and assume $\mathbf{m} \parallel \mathbf{x}$. In such a case, ${{\tilde k}_{1,2}} \approx  \mp \sqrt {2m^\ast \Delta}/\hbar$, 
$\tilde k{'_{1,2}} \approx  \mp i\sqrt {2m^\ast \Delta}/\hbar$ and we introduce the following dimensionless quantity $\epsilon  = \sqrt {(2m^\ast \alpha ^2)/(\hbar^2 \Delta)}$. Up to the lowest order in $\epsilon$, the barrier spinors can be written as ${{\tilde \chi }_{1,2}} = \left( {\begin{array}{*{20}{c}}
	1 \\ 
	{{e^{ \mp i\epsilon }}} 
	\end{array}} \right)$
for propagating states and ${{\tilde \chi }_1}' = \left( {\begin{array}{*{20}{c}}
	1 \\ 
	{ - 1 - \epsilon } 
	\end{array}} \right)$,
${{\tilde \chi }_2}' = \left( {\begin{array}{*{20}{c}}
	{ - 1 - \epsilon   } \\ 
	1 
	\end{array}} \right)$
for decaying states.
We can obtain the reflection coefficients in the limit of small $\epsilon$,
\begin{equation}\label{eq:interface_reflections}
\begin{gathered}
{r_{s1,s2}} = 1 - \left[\frac{{ \pm \left( {1 + {e^{i\beta }}} \right)\left( {u - 1} \right){k_1}}}{{\left( {{e^{i\beta }} - u} \right){k_0}}} \right. \quad\quad\quad\quad\quad \\
+ \left. \frac{{i\left( {{e^{i\beta }} - 1} \right)\left( {1 + u} \right){\left| {{k_2}} \right|}}}{{\left( {{e^{i\beta }} - u} \right){k_0}}} \pm i \right]\epsilon + O({\epsilon ^2}),  \\ 
\end{gathered} 
\end{equation}
with $k_0 \equiv 2m^\ast \alpha /\hbar ^2$.  Here, $u, \beta$ are defined in the spinors for the lead eigenstates related to the Eq.~(\ref{eq:2_lead_states}).

The total phase shift at the interfaces is given by
\begin{equation}
\begin{gathered}
{\varphi _2} = \frac{{{{\left( {1 + u} \right)}^2}\left| {{k_2}} \right|\left( {\cos \beta  - 1} \right) + \left( {1 - {u^2}} \right){k_1}\cos \beta }}{{\left( {1 + {u^2} - 2u\cos \beta } \right){k_0}}}  \\
\times 2\epsilon  + O({\epsilon ^2}).  \\ 
\end{gathered} 
\label{eq:reflection_phase_shift}
\end{equation}

The formation of standing waves in the barrier requires ${\varphi _1} + {\varphi _2} = 2n\pi $ with integer $n$, which gives the special resonance condition 
\begin{equation}\label{eq:special_resonance}
\left( {{{\tilde k}_2} - {{\tilde k}_1}} \right)d = 2n\pi  - \delta,
\end{equation}
with correction (up to the lowest order in $\epsilon$)
\begin{equation}
\delta  = \frac{{{{\left( {1 + u} \right)}^2}\left| {{k_2}} \right|\left( {\cos \beta  - 1} \right) + \left( {1 - {u^2}} \right){k_1}\cos \beta }}{{\left( {1 + {u^2} - 2u\cos \beta } \right){k_0}}}2\epsilon.
\label{eq:special_resonance1}
\end{equation}
The higher-order resonances in Fig.~\ref{fig:other_resonances}
occur at large proximity-induced exchange field,  
so the correction $\delta$ in Eq.~\eqref{eq:special_resonance1} is negligibly small. As a result, the resonance condition analogous to the spinless systems
\begin{equation}
\left( {{{\tilde k}_2} - {{\tilde k}_1}} \right)d = 2n\pi, \text{ with } n \text{ integer}
\label{eq:spinless_resonance}
\end{equation}
still works well in such cases.

\bibliography{TAMR}
\end{document}